\documentclass[journal]{IEEEtran}
\usepackage{amsmath}
\usepackage{amssymb} 
\usepackage{mathrsfs} 
\usepackage{bm} %bold in math condition
\usepackage{algorithm,color,soul}
\usepackage{graphicx,caption,subcaption}

\renewcommand{\thetable}{\arabic{table}}

\usepackage{booktabs} % setting tabular rules (\hline's) of different weight
\usepackage{ifpdf}
\usepackage{cite}
\usepackage{amsmath}
\usepackage{algorithmic}
\usepackage{array}

\ifCLASSOPTIONcaptionsoff
  \usepackage[nomarkers]{endfloat}
 \let\MYoriglatexcaption\caption
 \renewcommand{\caption}[2][\relax]{\MYoriglatexcaption[#2]{#2}}
\fi
\usepackage{url}
\newtheorem{Theorem}{Theorem}
\newtheorem{Definition}{Definition}

\newtheorem{Proof}{Proof}

\newtheorem{example}{Example}

\newtheorem{Property}{Property}

\newcommand{\ff}{\frac}                  
\newcommand{\f}{\dfrac}

\newcommand{\bb}{\mathbb}                    
\newcommand{\mb}{\mathbf}
      
\newcommand{\grad}{\triangledown}             
       
\newcommand{\z}{x}   
   
\newcommand{\x}{x}   
\newcommand{\Z}{z}
\newcommand{\impd}{q}
\newcommand{\Cor}{\Omega}
\newcommand{\C}{\mathcal{C}}   % cluster 
\newcommand{\argmin}[1]{\underset{#1}{\operatorname{arg}\operatorname{min}}\;}

\newcommand{\R}{\mathbb{R}}

\newcommand{\red}[1]{\textcolor{black}{{#1}}}

\newcommand{\blue}[1]{\textcolor{black}{{#1}}}
\newcommand{\cyan}[1]{\textcolor{black}{{#1}}}

\newcommand{\vertiii}[1]{{\left\vert\kern-0.25ex\left\vert\kern-0.25ex\left\vert #1 
		\right\vert\kern-0.25ex\right\vert\kern-0.25ex\right\vert}}
\newcommand{\verti}[1]{{\left\vert #1\right\vert}}

\begin{document}
\title{Cluster Prediction for Opinion Dynamics \\ from Partial Observations}
%Cluster prediction for opinion dynamics from partial observations


%\author{\IEEEauthorblockN{Zehong Zhang\IEEEauthorrefmark{1}, and
%		Fei Lu\IEEEauthorrefmark{1}}
%	\IEEEauthorblockA{\IEEEauthorrefmark{1}Department of Mathematics,
%		Johns Hopkins University, Baltimore, MD 21218 USA}
%	% \IEEEauthorblockA{\IEEEauthorrefmark{2}Department of Mathematics, Johns Hopkins University, Baltimore, MD 21218 USA}
%		}

\author{Zehong~Zhang,~and~Fei~Lu% <-this % stops a space	
	\thanks{Zehong Zhang and Fei Lu are with the Department of Mathematics,
		Johns Hopkins University, Baltimore, MD 21218, USA (e-mail: zzehong1@jhu.edu; feilu@math.jhu.edu).}}

%\thanks{Manuscript received December 1, 2012; revised August 26, 2015. 
%Corresponding author: M. Shell (email: %http://www.michaelshell.org/contact.html).}

%{Shell \MakeLowercase{\textit{et al.}}: Bare Demo of IEEEtran.cls for IEEE %Transactions on Magnetics Journals}




% for Transactions on Magnetics papers, we must declare the abstract and
% index terms PRIOR to the title within the \IEEEtitleabstractindextext
% IEEEtran command as these need to go into the title area created by

\maketitle

% \IEEEdisplaynontitleabstractindextext

% \fi
\IEEEpeerreviewmaketitle

\begin{abstract}
We present a Bayesian approach to predict the clustering of opinions for a system of interacting agents from partial observations. The Bayesian formulation overcomes the unobservability of the system and quantifies the uncertainty in the prediction.  We characterize the clustering by the posterior of the clusters' sizes and centers, and we represent the posterior by samples. To overcome the challenge in sampling the high-dimensional posterior, we introduce an auxiliary implicit sampling (AIS) algorithm using two-step observations. Numerical results show that the AIS algorithm leads to accurate predictions of the sizes and centers for the leading clusters, in both cases of noiseless and noisy observations. In particular, the centers are predicted with high success rates, but the sizes exhibit a considerable uncertainty that is sensitive to observation noise and the observation ratio.  
\end{abstract}


\section{Introduction}
\IEEEPARstart{C}{lustering} behavior in a network of interacting agents or particles arises in a vast range of disciplines \cite{Krause2000,VZ2012,motsch2014heterophilious}. In the context of opinion dynamics of social networks, local interactions among agents cause opinions to evolve, formulating one or more clusters of opinions. While the striking phenomenon of consensus (one cluster) has attracted long-standing interest, non-consensus clustering, in which multiple stable clusters coexist, has attracted increasing interest to resemble the real-life social network \cite{shao2009_DynamicOpinion,li2013_NonconsensusOpinion,proskurnikov2016_OpinionDynamics,lancichinetti2012_ConsensusClustering}. 
Such clustering of opinions or communities have a profound impact on the network, so it is of great importance to predict these clusters from observations, which are often partial, at an early stage.

We investigate the prediction of clusters for multi-agent opinion dynamics with multiple clusters, from short-time partial observations which may be contaminated by white noise. In particular, our objective is to predict the sizes and centers of the leading clusters. We assume the system is known (we refer to \cite{BFHM17,LZTM19,LMT19} and the references therein for the learning of the governing equation from data). To predict the clustering, one may estimate all agents' current opinions and use them as an initial configuration for prediction. However, we show that it is an ill-posed inverse problem to estimate the current state from partial observations (widely-studied as observability in control, see e.g., \cite{tucsnak2009observation}). We propose a Bayesian formulation to make the problem well-posed: we estimate the posterior distribution of the states conditional on the observations. We represent the posterior by samples, which provide initial configurations for prediction.  This procedure yields a posterior for the clusters' sizes and centers, quantifying the uncertainty in prediction. 
% In this paper, based on given observation information, which is not rich enough, our objective is to predict the 'type' of clustering, i.e. the size and center of each cluster. Since the limitation of observation, we need to address this problem within perspective of Bayesian Inference. 

The major challenge in the Bayesian approach is to generate samples for the high-dimensional posterior. Due to the intrinsic symmetry of the nonlinear opinion dynamics, the non-Gaussian posterior has multiple local extrema, which posed a hurdle for the performance of Sequential Monte Carlo (SMC) methods\cite{doucet2009tutorial,gilks2001following,berzuini2001resample}, including the optimal (one-step-observation) importance sampling methods such as implicit sampling \cite{CT09}. The symmetry and weak correlation between the states also prevent the feedback control or nudging methods \cite{lunasin2017_FiniteDetermining,foias2016discrete,liu2015discontinuous,nijmeijer2001dynamical,auroux2005back,zhu2016implicit,LWM19} based on dominating modes in the observation. 

We overcome the challenge by introducing an Auxiliary Implicit Sampling (AIS) algorithm that makes use of two-step observations, which is a sequential Monte Carlo method that combines the ideas from auxiliary particle filters \cite{pitt1999filtering}, implicit sampling\cite{CT09} and feedback control \cite{auroux2005back}. We also introduce an MCMC-move step to reduce sample degeneracy and an information move step to reject non-physical samples.  

Numerical tests show that our AIS algorithm leads to accurate prediction of the sizes and centers for the leading clusters, in both cases of noiseless and noisy observations. In particular, the centers of the leading clusters are predicted with a high success rate, but the size of the leading cluster exhibits a considerable uncertainty that is sensitive to observation noise and the observation ratio. Our AIS algorithm brings improvement to implicit sampling, and both outperform the sequential importance sampling with resampling (SIR) method.

\blue{Our AIS algorithm is applicable to general state-space models with Gaussian noises and linear observation models, particularly to those with symmetric and weakly correlated state variables. It derives effective importance densities using two-step observations. It supplies an efficient SMC component to the algorithms that combine SMC with MCMC, such as the particle MCMC methods or the nested particle filters  \cite{andrieu2010particle,crisan2018nested,lindsten2014particle}. Our importance densities also provide effective candidates for algorithms with multiple importance densities \cite{liu2000multiple,pandolfi2010generalization,martino2018group}. }

The exposition in our manuscript proceeds as follows. In Section \ref{sec:Bayes}, we define clusters for opinion dynamics with local interactions, prove that the inverse problem of state estimation from partial observation is ill-posed, and propose a Bayesian formulation for cluster prediction. To represent the posterior, we introduce in Section \ref{sec:Method} an auxiliary implicit sampling algorithm that designs importance densities based on two-step observations. Section \ref{Simulation} examines the performance of the AIS algorithm in numerical simulations.  Finally, Section \ref{sec:conclusion} concludes the paper with discussions.

\section{Bayesian approach to cluster prediction} \label{sec:Bayes} 
Consider a group of $N$ agents, each with an opinion at time $t$ quantified by $\x_t^i \in \R^d$, interacting with each other according to a first-order difference system: 
\begin{equation}  \label{opinion dynamic}
x_{t+1}^i = x_{t}^i + \f \alpha N \mathop \sum_{j=1}^N \phi(\|\x_{t}^j-\x_{t}^i\|)(\x_{t}^j-\x_{t}^i). 
\end{equation}
Here, the positive constant $\alpha$ is a scaling parameter and the interaction kernel $\phi$ is a non-negative function supported on $[0,R]$. 
The agents interact locally, only with those opinions that are ``close'' in the sense that the pairwise distance $\|\x_{t}^i-\x_{t}^j\|$ is less than $R$.

\cyan{Our goal is to predict the clustering of the opinion dynamics, particularly the sizes and the centers of the leading clusters, from partial data. The data consists of trajectories of partial agents for a relatively short time, far before the system forms clusters. To quantify the uncertainty due to the random initial condition and the measurement error in data (which we assume to be Gaussian), we present a Bayesian approach.  More specifically, we would like to numerically approximate the posteriors of the sizes and the centers of the largest clusters in the steady-state of the system (see Eq.\eqref{eq:postCluster} for a precise description). 
}

In this section, we provide a quantitative definition for clustering and discuss clustering prediction from partial observations. We show that it is an ill-posed inverse problem to predict the clustering by estimating all agents' trajectories. We introduce a Bayesian approach to make the problem well-posed, providing a probabilistic quantification of the uncertainty in the prediction.  

%%%%%%%%%======== 
\subsection{Definition of clusters} \label{Doci}
Due to the local interaction between agents, clusters of opinions will emerge, in which each agent only interacts with agents within the same cluster. More precisely, we define the system is in a clustered status as follows: 
\begin{Definition}[Clustered status] Let $\x_{t} \in \bb R^{dN}$ be the state of the system \eqref{opinion dynamic} with a local interaction kernel $\phi$ supported on $[0,R]$. We say the system is \textbf{clustered} if the index set $\{1,2,\ldots,N\}$ of agents can be partitioned into disjoint clusters  $ \C_1(t),...,\C_m(t) $ such that for any $ i \in \C_{k_1}(t) $ and $j \in \C_{k_2}(t)$:
	\begin{enumerate}
		\item[(i)] if $k_1 = k_2$, then $\|\x_t^i - \x_t^j\| < R $,
		\item[(ii)] if $k_1 \neq k_2$, then $ \|\x_t^i - \x_t^j\| > R $. 
	\end{enumerate}
\end{Definition}

An essential feature of the clustered status is that it is invariant in time: a clustered system will remain clustered with the same clusters. In particular, each cluster is isolated from other clusters; in each cluster, the agents formulate self-contained dynamics and concentrate towards a local consensus, the center of the cluster, since the interaction is symmetric (we refer to \cite{motsch2014heterophilious} for detailed discussions on clustering for local interactions). We summary this invariant feature as a property of the system. 
\begin{Property}[Invariants of a clustered system] \label{position and size}
	Suppose that at time $t_c$, the system \eqref{opinion dynamic} is clustered into $\{ \C_1,...,\C_K  \}$. Then, the system will remain clustered with the same clusters for all  $t \geq t_c$. In particular, the  sizes and the centers of the clusters are invariant in time: for all $t \geq t_c$, 
	\begin{equation}
	\begin{aligned}
	|\C_k| :&= |\C_k(t)| = |\C_k(t_c)|,  
	\\
	\overline{\x}_{\C_k} :& =  \f 1 {|\C_k(t)|} \mathop{\sum}\limits_{i \in \C_k(t)} \x_{t}^i = \f 1 {|\C_k(t)|} \mathop{\sum}\limits_{i \in \C_k(t)} \x_{t_c}^i . 
	\end{aligned}
	\end{equation}
	for each $k=1,\ldots, K$, where $ |\C_k| $ and $ \overline{\x}_{\C_k} $ denote the size (number of agents) and center (mean opinion of agents) of cluster $\C_k$, respectively.
\end{Property}
These invariants characterize the clustering (the large time behavior) of the opinion system. Therefore, our goal of clustering prediction is to estimate these invariants: the sizes and centers of the clusters, particularly those of the largest clusters.

	\subsection{Cluster identification from partial observations} \label{sec:Cifpod}
In practice, it is often the case that we can only observe or track partial of the agents. 
% This prompts the problem of cluster identification and prediction from partial observations. 
We consider the case that $N_1$ out of the $N$ agents are observed, with $ \Z_{1:T} \in \bb R^{TdN_1} $ denoting their trajectories. We will consider either noiseless or noisy observations. 
The original model \eqref{opinion dynamic} with initial distribution $\mu$, together with an observation equation, can be written as the following state space model: 
\begin{equation} \label{state-space model}% {SSM0}
\left\lbrace \begin{aligned}
\x_{t+1} & = g(\x_t),  \ \ \z_1 \sim \mu (\cdot),
\\
\Z_t & = H \x_t  + \xi_t,
\end{aligned} \right.
\end{equation} 
where $ g(\x_t) $ is the right-hand-side of \eqref{opinion dynamic}, and $ H: \bb R^{dN} \to \bb R^{dN_1} $ is a projection operator mapping the vector of opinions of all agents to its observed part, and $\xi_t$ are independent identical distributed (i.i.d.) Gaussian with distribution \red{$\mathcal{N}(0, \sigma_\xi^2 I_{dN_1})$ (with $\sigma_\xi=0$ if the observations are noiseless)}.

 Without lost of generality, we assume that the first $N_1$ agents are observed. For simplicity of notation, we denote   $H\x=(\x^1,...,\x^{N_1}) \in \bb R^{dN_1}$  with $ H= [ I_{dN_1}\mid 0 \times I_{dN_2} ] $ and with $ H_i\x = \x^i $ as the $i$-th observed agent. Similarly, for the unobserved agents, we define projection operator $ G: \bb R^{dN} \to \bb R^{dN_2} $ from the state $\x$ to its unobserved part, denoting $G\x=(\x^{N_1+1},...,\x^{N}) \in \bb R^{dN_2}$ with $ G= [ 0 \times I_{dN_1}\mid  I_{dN_2} ] $ and with $ G_i\x = \x^{N_1+i} $ as the $i$-th unobserved agent. We summarize the notation in Table \ref{tab:notation}. 
\begin{table}[!t] 
	\begin{center} 
		\caption{ \\ \text{Notation of variables in the state-space model}} \label{tab:notation}
		\begin{tabular}{ l  l }
		\toprule % \hline
			Notation   &  Description \\  \hline
			$\x=(\x^1,...,\x^N) \in \bb R^{dN}$            & state variable of the system            \\
			$H\x=(\x^1,...,\x^{N_1}) $, $H_i\x = \x^i$        & opinions of observed agents                       \\ 
			$G\x=(\x^{N_1+1},...,\x^{N})$ &  opinions of unobserved agents        \\
			$|\C_i|$ and $\overline{\x}_{\C_i}$                & size and center of cluster $\C_i$ \\
			$\x_{1:t}=(\x_1,...,\x_t) \in \bb R^{tdN}$   & trajectory of all agents   \\
			$\Z_{1:t}=(\x_1,...,\x_t) \in \bb R^{tdN_1}$   & trajectory of observed agents  \\
			\bottomrule	
		\end{tabular}  
	\end{center}
\end{table}

To predict the clustering, which is the large time behavior of the dynamics, based on observations up to time $T$, a natural idea is to (i) estimate the state of the system at time $T$, and (ii) use the estimated state as an initial condition for a long time simulation until the system is clustered. 
For Step (i), one may wish to find a trajectory of the state variable that fits the observation data. However, the following section shows that even with noiseless partial observations, it is an ill-posed inverse problem to identify the trajectory $\x_{1:T}$ from observation $\Z_{1:T}$. Also, whereas a regularization can make the problem well-posed in a variational approach, it leads to a challenging high-dimensional optimization problem on the path space and there may be many local minima caused by the symmetry of the system. 
Instead, we adopt a Bayesian approach that avoids high-dimensional optimization and quantifies the uncertainty in prediction.

\subsection{State estimation and observability} \label{Observability}
In general, it is an ill-posed inverse problem to estimate the trajectory of all agents from partial noiseless observations. We demonstrate this by an example of symmetric trajectories and by proving that the unobserved trajectories can not be uniquely determined in linear systems, referred to as unobservability in control (see e.g., \cite{tucsnak2009observation}), when more than one agents are unobserved.  

\begin{figure}[!t]
	\centering
	\begin{subfigure}[b]{0.48\linewidth}
		\includegraphics[width=1\linewidth]{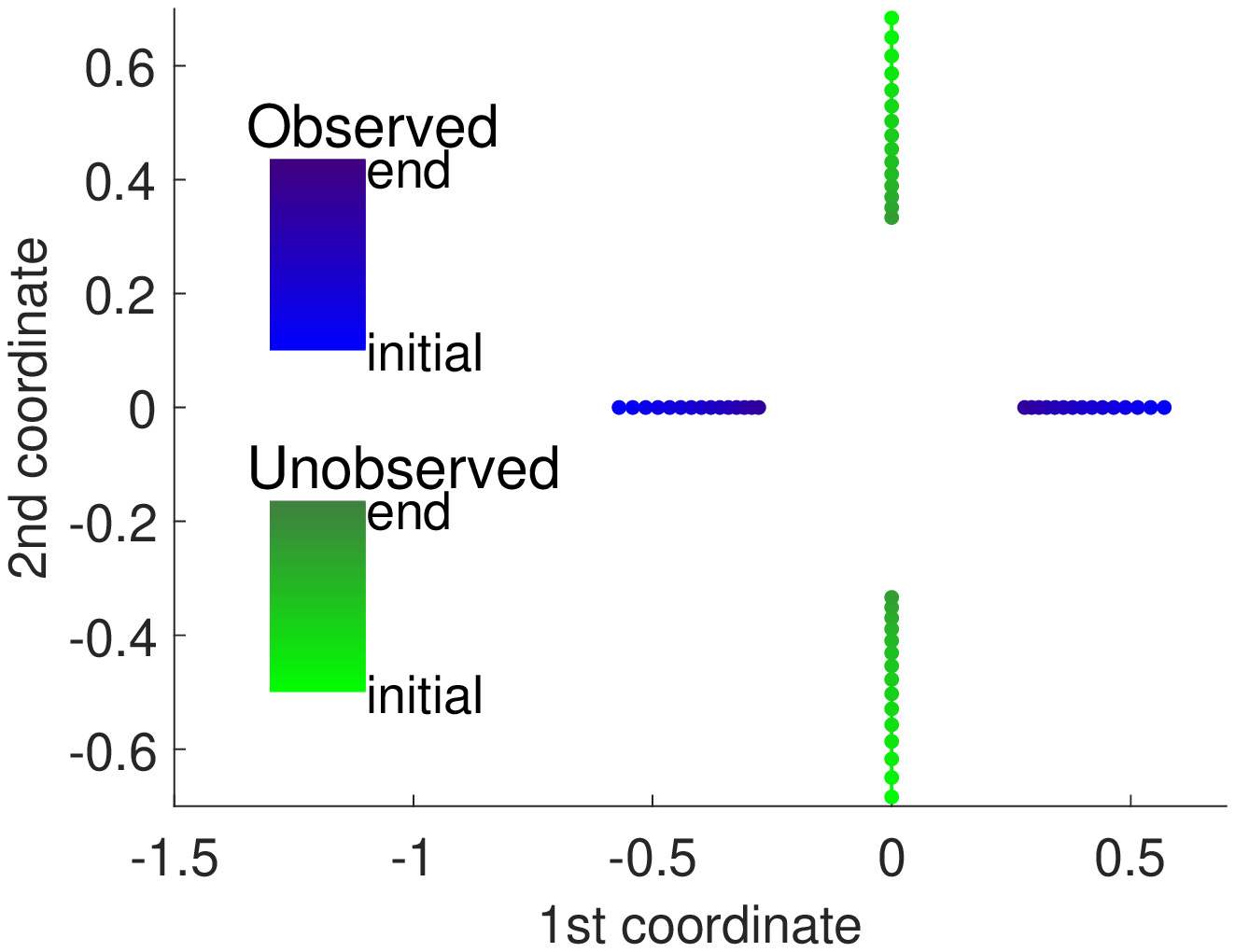}
	\end{subfigure}
\hfil
	\begin{subfigure}[b]{0.48\linewidth}
		\includegraphics[width=1\linewidth]{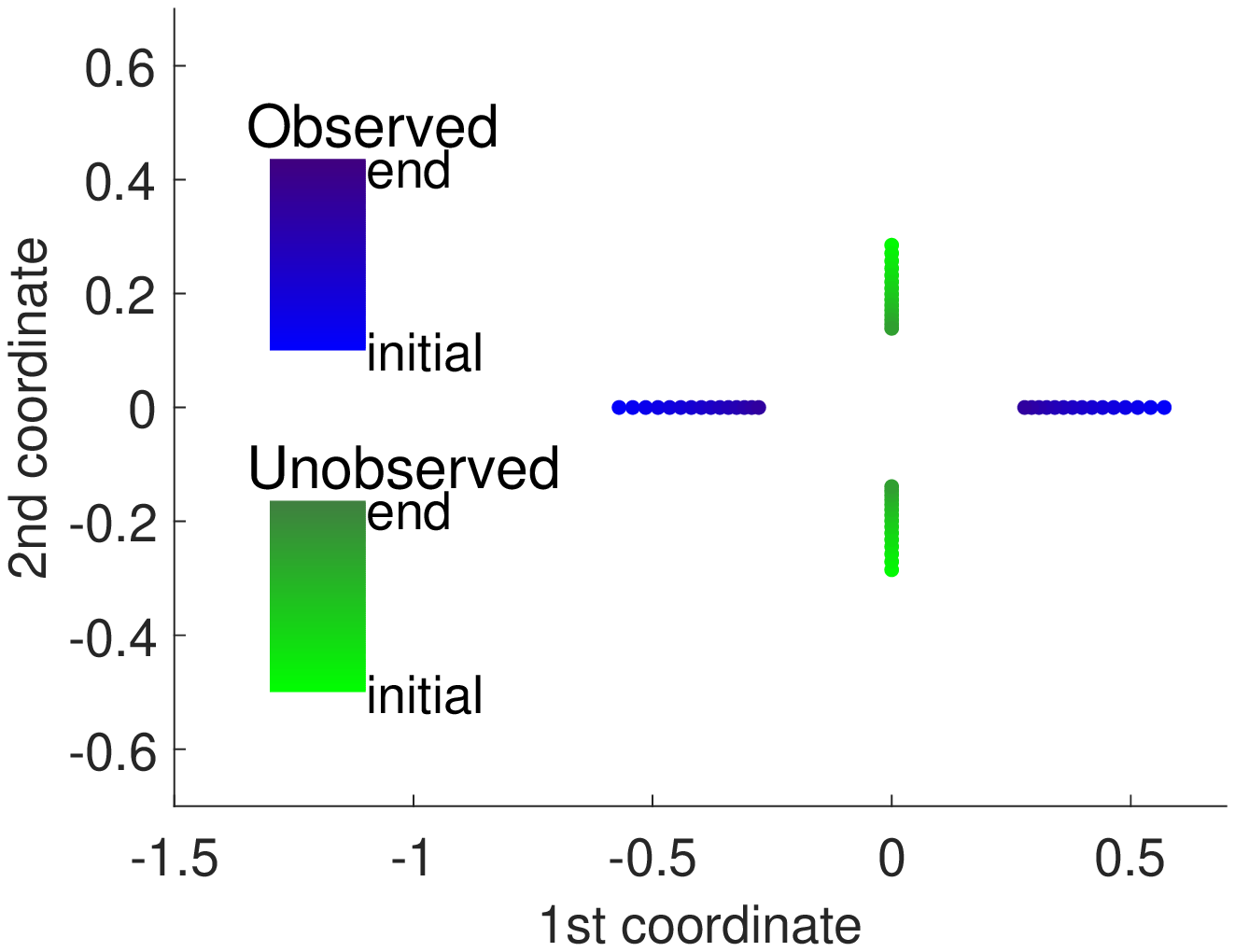}
	\end{subfigure}\hspace{-0.09in}
\hfil
	\caption{Illustration of symmetric trajectories: same observed trajectories (blue points) are generated from different configurations (with different unobserved trajectories in green). The color changes from light to dark to indicate time increasing from initial to end-time of observation.}
	\label{special unobservable cases} 
\end{figure}

The next example shows that as long as more than two agents are unobserved, there could be symmetric trajectories, making it an ill-posed problem to identify the trajectories.   
\begin{example}[Symmetric trajectories]
	Consider a system with $N=4$ agents in $\R^2$ and suppose that we observe $N_1=2$ of them. Figure \ref{special unobservable cases} illustrates that two different configurations can lead to the same observations. The symmetric positions of the two unobserved agents canceled out their different influence on the observed agents. 
\end{example}

%\begin{example}[Linear system: observability requires $N_1\geq N-1$]  \end{example}

%In general, for a system with a piecewise constant local interaction function (e.g. the agents interact only with agents with similar opinions), if all opinions interact with others, we may view the system as a linear system with global interaction. 

The following theorem show that it is an ill-posed problem to estimate the states of the system when more than one agents is unobserved in the case of linear systems. 
 \begin{Theorem} [Observability for linear opinion dynamics] Consider the linear dynamics with $\phi\equiv 1$ in \eqref{opinion dynamic}, and suppose that we observed the trajectory of $N_1$ agents. Then, the trajectories of the unobserved agents can be uniquely determined if and only if 
%	Partially observed linear opinion dynamics are observable if and only if
$	N_1 \geq N-1.$
	
%	As a consequence, to identify the trajectory  $\x_{1:T}$ from the noiseless observations $z_{1:T}$, the number of observed agents must be $N_1\geq N-1$. 
	\end{Theorem}
    \begin{Proof}
%    Suppose that the system evolves as a linear system during the time interval of observation $[0,T]$. Then, for $0\leq t \leq T$, 
We only need to consider $ N_1 \leq N-1$. We can write the system as  
	\[
	\left\lbrace \begin{aligned} 
	\x_{t+1} & = \alpha A \x_t + \x_t ,  	\\
	\Z_t & = H \x_t, 
	\end{aligned} \right.
	\]
	where $ A \in \bb R^{dN} \times \bb R^{dN} $ is a constant matrix, 
	\begin{align*}
	A =\left(\begin{array}{cccc} 
	c_1I_d& c_2I_d&\cdots & c_2 I_d  \\ 
	c_2 I_d & c_1I_d&\cdots& c_2 I_d \\ 
	\vdots & \vdots & \ddots & \vdots\\ 
	c_2 I_d& c_2 I_d&\cdots & c_1I_d
	\end{array}  \right)
	\end{align*}
	with $c_1 = - \frac{(N-1)}{N}$ and $c_2 = \frac{ 1 }{N}$. % By the observability theory \cite{tucsnak2009observation}, we present the following theorem for this partially observed linear opinion dynamics. 
By the observability theory \cite{tucsnak2009observation}, %Linear observability theory shows us that the trajectory 
 the trajectory $\x_{1:T}$ can be uniquely determined from the observations $z_{1:T}$ if and only if $ \mathrm{rank}\ (W) = dN $, where
    \[
    W:= \left[ H^\intercal\mid A^\intercal H^\intercal \mid ... \mid (A^\intercal)^{n-1}H^\intercal \right].
    \]
 To compute $\mathrm{rank}(W)$, note that  $A^\intercal=A$ and $ A = Q \Lambda Q^\intercal$, where $ \Lambda = \mathrm{diag}(-I_{d(N-1)}, 0\times I_d) $ and $Q$ is a unitary matrix. Recalling that $ H= [ I_{dN_1}\mid 0 \times I_{dN_2} ] $, we have $  (A^\intercal)^k H^\intercal = (-1)^{k-1}AH^\intercal $ for $k=1,\ldots,n-1$. Thus,  
    \begin{align*}
    \mathrm{rank}(W) =\mathrm{rank}( [ H^\intercal \mid AH^\intercal ] ) =  (N_1 + 1) \times d. 
    \end{align*}
 
    \end{Proof}

\subsection{Bayesian estimation of states and clusters} \label{Beosac}
In a Bayesian approach, we view the states and the invariants of the clusters as random variables and we aim to represent their posteriors conditional on the observations. 

\iffalse
We begin by writing the system in the form of a state space model: 
 \begin{equation} \label{state-space model}
 \left\lbrace \begin{aligned}
\x_{t+1} & = g(\x_t) \ \ \ \z_1 \sim \mu (\cdot),
\\
\Z_t & = H\z_t + \xi_t,
\end{aligned} \right.
\end{equation}
where % the state-noise $ \epsilon_t $ is zero in the true model, and 
\hl{the observation-noise $ \xi_t $ are independent identical distributed (i.i.d.) Gaussian with distribution $\mathcal{N}(0,\sigma_\xi^2)$( we set $\sigma_\xi=0$ in the noiseless case).} 
\fi
Recall the state space mdoel in \eqref{state-space model}. 
When observation is noise free, the randomness of the states comes from the initial distribution $\mu$. Conditional on observations $\Z_{1:T}$, we denote by $p(|\C_i| \mid \Z_{1:T})$, and $p(\overline \x_{\C_i} \mid \Z_{1:T})$  the posteriors of the size and center of cluster $\C_i$, and similarly the posterior of the state variables, as in Table \ref{tab:notationBayes}. 

\begin{table}[!t]
	\setlength{\abovecaptionskip}{0pt}
	\begin{center}
		\caption{ %\\ Notation of variables in Bayesian approach, conditional on observations $\Z_{1:T}$.}   \label{tab:notationBayes}
		\\ \text{Notation of
				variables in the Bayesian approach }}  \label{tab:notationBayes}
		%	\textsc{conditional on observations} $\Z_{1:T}$.
		%	\\ \vspace{5pt}
		\begin{tabular}{ l  l }
		\toprule 
		Notation  & Description \\ \hline
			$p(\x_{1:T}\mid \Z_{1:T})$,              & posterior of $\x_{1:T}$ conditional on  $ \Z_{1:T}$     \\
			$\widehat p(\x_{1:T}\mid \Z_{1:T})$ & empirical approximation of    $p(\x_{1:T}\mid \Z_{1:T})$                    \\ 
			$\{ \x_{1:t}^{(s)}, w_t^{(s)}\}$            & samples and weights 	 \\
			  $p(|\C_i| \mid \Z_{1:T})$, $p(\overline \x_{\C_i} \mid \Z_{1:T})$  & posteriors of $|\C_i|$ and $\overline \x_{\C_i}$ \\
			  \bottomrule	
		\end{tabular}  
	\end{center}
\end{table}

These posteriors of the invariants depend on the initial distribution as well as the system, and can not be expressed analytically in general. They depend on the posterior of the state $p(\x_{1:T}\mid \Z_{1:T})$, particularly $p(\x_{T}\mid \Z_{1:T})$. % when the opinion dynamics is deterministic. 
% The analytical expression of these posteriors of the state variables may be available, but 
They are high-dimensional and non-Gaussian. 

We approximate these distributions by Monte-Carlo methods: we draw a set of weighted samples  (with normalized weights),  $\{ \x_{1:t}^{(s)}, w_t^{(s)} \}_{s \in \{ 1,...,S \}}$, by a sequential Monte Carlo method (to be introduced in the next section) from the target distribution $p(\x_{1:T}\mid \Z_{1:T})$, and obtain empirical approximations of these distributions. For instance, the posterior $p(\x_{T}\mid \Z_{1:T})$ is approximated by 
\[
\widehat p(\z_T \mid \Z_{1:T}) = \sum_{s=1}^S  w_T^{(s)} \delta_{\z_{T}^{(s)}} (\z).
\]

By running the original system from each of the samples $\{\x_T^{(s)}\}$ until the status of clustered,  we obtain weighted samples for the invariance of clusters $\{ \overline{\x}_{\C_i}^{(s)} , w_T^{(s)} \}_{s \in \{ 1,...,S \}}$ and $\{ |\C_i^{(s)}| , w_T^{(s)} \}_{s \in \{ 1,...,S \}}$.   
With these weighted samples, we have the empirical posterior to quantify the uncertainty in cluster prediction:
\begin{equation}\label{eq:postCluster}
\left\lbrace \begin{aligned}
\widehat  p(\overline{\x}_{\C_i} \mid \Z_{1:T})  &=  \sum_{s=1}^S  w_T^{(s)} \delta_{\overline{\x}_{\C_i}^{(s)}} (\overline{\x}_{\C_i}), 	\\
\widehat  p(|\C_i| \mid \Z_{1:T}) & =  \sum_{s=1}^S  w_T^{(s)} \delta_{ |\C_i^{(s)}| } ( |\C_i|). 	\end{aligned} \right. 
\end{equation}

With the weighted samples, we can efficiently approximate the statistics by the samples. For example, the expectations of the size and center of cluster $\C_i$ are  
\begin{equation}\label{eq:mean}
\begin{aligned}
\bb E (\overline{\x}_{\C_i}) \approx \quad & \widehat{\overline{\x}_{\C_i}} := \sum_{s=1}^S \overline{\x}^{(s)}_{\C_i} \cdot w_{T}^{(s)}, 	\\
\bb E ( |\C_i| ) \approx \quad & \widehat{|\C_i|} := \sum_{s=1}^S |\C_{i}^{(s)}| \cdot w_{T}^{(s)}. 
\end{aligned}  
\end{equation}

%%%%%%%%%%%%%%%%%%%%%%=============
%%%%%%%%%%%%%%%%%%%%%%=============
%%%%%%%%%%%%%%%%%%%%%%=============	
\section{Sampling the posterior} \label{sec:Method}
To initiate the ensemble simulation for prediction, we draw samples from the conditional distribution of the current state, $ p(\z_T \mid \Z_{1:T}) $, which is the marginal distribution of the posterior distribution $ p(\z_{1:T}\mid \Z_{1:T}) $. This posterior is high-dimensional, nonlinear and non-Gaussian, therefore it is difficult to sample directly, even when its analytical form is explicitly available.

We will adopt a Sequential Monte Carlo (SMC) strategy (we refer to \cite{doucet2009tutorial} for a review), with a combination of implicit sampling \cite{CT09} and Auxiliary particle filtering, and some specialized MCMC-move and information-move.

To avoid degenerate distributions, we introduce artificial noises to the state-space model \eqref{state-space model} from section \ref{Beosac} 
\[
\left\lbrace \begin{aligned}
\z_{t+1} & = g(\z_t) + \epsilon_t, \ \ \ \z_1 \sim \mu (\cdot),
\\
\Z_t & = H\z_t + \xi_t.
\end{aligned} \right.
\]
where % $\epsilon_t$ and $\xi_t$ are assumed to be i.i.d.~centered Gaussian, 
 \red{$ \epsilon_t \sim \mathcal{N}(0, \sigma_\epsilon^2 I_{dN})$ and $ \xi_t \sim \mathcal{N}(0, \sigma_\xi^2 I_{dN_1}) $ with $\sigma_\epsilon>0$ and $\sigma_\xi>0$. } \red{In particular, we set the variances so that (i) the artificial noises are relatively small with respect to the signal; (ii) the important densities (to be introduced below in our sequential Monte Carlo algorithm) have centers relying on the state model more than the observations and they have relatively large variances to explore large ranges. In view of the importance densities in \eqref {MAP_point}--\eqref{Implicit_Sampling} and \eqref{eq:yAb}--\eqref{importanceDensity}, we will set $\sigma_\xi/ \sigma_\epsilon <1$.   
}
Here we assume the variances to be constants for simplicity, but they can vary in time to improve the algorithm.

% In the original model, these noises are zero,  leading to degenerate distributions that often result in deficient degenerate samples in SMC algorithms. We set the noise  $\epsilon_t$ in the state model at a level that enables the particles to explore possible states but is negligible with respect to the interaction force. As for observation noise $\xi_t$, we set it at a level that XXX balance between degeneracy speed and weighting different samples.XXX

\subsection{Sequential Monte Carlo sampling}

The SMC methods, or particle filters, are a set of sequential importance sampling algorithms that approximates the high dimensional distribution  $ p(\z_{1:t}\mid \Z_{1:t}) $ by its empirical distribution from weighted samples $\{ \z_{1:t}^{(s)}, w_t^{(s)} \}_{s \in \{ 1,...,S \}}$: 
\[
\widehat p(\z_{1:t}\mid \Z_{1:t}) := \f 1 { \sum_{s=1}^S w_t^{(s)}} \sum_{s=1}^S  w_t^{(s)} \delta_{\z_{1:t}^{(s)}} (\z), 
\]
where $\delta$ is Dirac delta mass. The samples $\{\z_{1:t}^{(s)}\}$ are drawn from an importance distribution $q(\x_{1:t} \mid \Z_{1:t}) $ and the weights are computed from 
\begin{equation}\label{eq:wt_seq}
w(x_{1:t}\mid \Z_{1:t}) = \f {p(\z_{1:t} \mid \Z_{1:t})} {q(\z_{1:t} \mid \Z_{1:t})}.
\end{equation}
The key idea of SMC is to generate the weighted samples sequentially from a recursive importance density, 
\begin{equation}
q(\z_{1:t} \mid \Z_{1:t}) = q(\z_1) \prod_{k=2}^t q(\z_k \mid \z_{1:k-1}, \Z_{1:k} ),
\end{equation}
which is constructed based on the recursive representation of the posterior distribution:
\begin{equation} \label{eq:rec-Imp}
p(\x_{1:t} \mid \Z_{1:t}) = p(\x_{1:t-1} \mid \Z_{1:t-1})  \f { p(\x_t\mid\x_{t-1}) p(\Z_t \mid \z_t) } {p(\Z_t \mid\Z_{1:t-1})},
\end{equation}
That is, at time $t$, conditional on previous samples $\{ \z_{1:t-1}^{(s)}, w_{t-1}^{(s)} \}_{s \in \{ 1,...,S \}}$, one generates weighted samples $\{\z_t^{(s)} \}$ from importance densities  $\{ q(\z_t \mid \z_{1:t-1}^{(s)}, \Z_{1:t} ) \}$ and compute their weights by 
\begin{equation}\label{eq:weight}
w_t^{(s)} = w_{t-1}^{(s)} \cdot \f { p(\Z_t \mid \z_t^{(s)}) \cdot p(\z_t^{(s)} \mid \z_{t-1}^{(s)}) } {q(\z_t^{(s)} \mid \z_{1:t-1}^{(s)}, \Z_{1:t} )}.
\end{equation}

Clearly, the above weight  $w_t^{(s)}$ is proportional to the analytical weight $w(x_{1:t}^{(s)}\mid \Z_{1:t})$ since $p(\z_{1:t}^{(s)} \mid \Z_{1:t})\propto p(\x_{1:t-1}^{(s)} \mid \Z_{1:t-1}) \cdot  p(\x_t^{(s)}\mid\x_{t-1}^{(s)}) p(\Z_t \mid \z_t^{(s)}) $ and $ q(\x_{1:t}^{(s)} \mid \Z_{1:t-1}) = q(\x_{1:t-1}^{(s)} \mid \Z_{1:t-1})  \cdot q(\z_{t}^{(s)} \mid \z_{1:t-1}^{(s)}, \Z_{1:t}) $. 

Due to the recursive computation in \eqref{eq:weight}, all but a few of the weights will be almost zero as $t$ increases, and this is called sample degeneracy \cite{doucet2009tutorial}. As a result, the variance of our estimation $ \{\z_{t}^{(s)} \} $ may increase exponentially with $t$ (see e.g. \cite{kong1994sequential}). Resampling techniques are widely used to reduce the sample degeneracy by replacing low-weighted samples with high-weighted samples through resampling. \blue{A common strategy is to measure the sample degeneracy by effective sample size (ESS) \cite{kong1992note,liu2008monte,martino2017effective} and set a threshold for resampling: if the ESS falls below a threshold (typically $\frac{S}{2}$ or $\frac{2S}{3}$), then one resamples. In our study, we use the ESS defined by $
\mathrm{ESS}_t =  (\sum_{i=1}^S w_t^{(i)})^2 / (\sum_{i=1}^S (w_t^{(i)})^2) $ in \cite{liu2008monte}. }We use the resampling algorithm in \cite{kitagawa1996monte}, i.e., sample $ u$ from the uniform distribution $ \mathcal{U}([0,\ff 1 {S}]) $ and define a set of real number $ \{ U_j:= u + \f {j-1} {S} \}_{j=1,...,S} $. Then count the number of the set $ \{ U_j \mid \f {\sum_{i=1}^{i'-1} w_t^{(i)} } {\sum_{i=1}^{s} w_t^{(i)} } \leq U_j \leq \f  {\sum_{i=1}^{i'} w_t^{(i)} }  {\sum_{i=1}^{s} w_t^{(i)} } \}$ as the number of ``children'' of sample $\x^{(i')}$.

The essential of SMC methods is the design of importance densities, so that all samples have (almost) equal weights in each recursive step while staying on the trajectories with high likelihood. The algorithm based on a simple choice of $q(\z_t \mid \z_{1:t-1}, \Z_{1:t} ) = p(\z_t\mid \z_{t-1})$, often referred as sequential importance sampling with resampling (SIR), performs poorly (see section section \ref{Cp1}). 
Inspired by the ideas of implicit sampling (see Section \ref{NIs}) and Auxiliary particle filtering, we propose to construct Gaussian importance densities by a combination of them (see Section \ref{NIs}). To rejuvenate the samples, we will also introduce MCMC-move and information-move algorithms, which will be discussed in Section \ref{Mmove} and \ref{Imove}, respectively.

\subsection{Optimal one-step importance sampling} 

The one-step optimal importance density is  %that combines the prior from the state model and the likelihood of observation
\begin{equation} \label{optimal important distribution}
q^{\mathrm{opt}}(\z_t \mid \z_{1:t-1}, \Z_{1:t} ) = \f { p(\Z_t \mid \z_t) \cdot p(\z_t \mid \z_{t-1}) } {p( \Z_{t} \mid \z_{t-1} )}.
\end{equation}
\blue{It is optimal because it is exactly the one-step posterior density, leading to uniform incremental weights in \eqref{eq:wt_seq}.} 

\blue{The density $q^{\mathrm{opt}}$ is Gaussian and can be sampled directly, because the observation model is linear and the noises $\epsilon_t$ and $\xi_t$ in the state-space model \eqref{state-space model} are Gaussian. } In general nonlinear non-Gaussian cases, it is difficult to draw samples from $q^{\mathrm{opt}}$ directly, and one may resort to implicit sampling in \cite{CT09} or  \cite{morzfeld2015parameter} to draw samples in the high probability region. 

To sample $q^{\mathrm{opt}}$, we need only its mean and covariance, which are the maximum a posteriori (MAP) and the Hessian of the negative logarithm of the posterior, respectively. More specifically, %since $ p( \Z_{t} \mid \z_{t-1} ) = \int p(\Z_t \mid \z_t) p(\z_t \mid \z_{t-1})d\z_t$, we can generate samples of $q^{\text{opt}}(\z_t \mid \z_{1:t-1}, \Z_{1:t} )$ by drawing weighted samples from $p(\Z_t \mid \z_t) p(\z_t \mid \z_{t-1})$, 
%\begin{align*}
%q^{\text{opt}}(\z_t \mid \z_{1:t-1}, \Z_{1:t} ) & = \f { p(\Z_t \mid \z_t) \cdot p(\z_t \mid \z_{t-1}) } {p( \Z_{t} \mid \z_{t-1} )} \\ 
%& \propto p(\Z_t \mid \z_t) p(\z_t \mid \z_{t-1}) ,
% \end{align*}
we compute the minimizer and Hessian of the negative log function of $ p(\Z_t \mid x) p(\x \mid \z_{t-1}) $: 
\[
F(\x) =  \f {(\Z_t - H\x)^2} {2\sigma_\xi^2} + \f{(\x - g(\x_{t-1}))^2} {2 \sigma_\epsilon^2}.
\]
This function is quadratic and its minimizer is 
\begin{align} \label{MAP_point}
\begin{aligned}
\x^*_t % & = \begin{bmatrix} H g(\x_{t-1})  +  \alpha_t (\Z_t - Hg(\x_{t-1})) \\ Gg(\x_{t-1}) \end{bmatrix} \\
& = g(\x_{t-1}) + \lambda_t^* \begin{bmatrix}\mb I_{dN_1\times dN_1} \\ \mb 0_{dN_2\times dN_1} \end{bmatrix}  (\Z_t - Hg(\x_{t-1}))
\end{aligned}
\end{align}
% H\x^*_t % & = \f { \sigma_\epsilon^2 \Z_t + \sigma_\xi^2 Hg(\x_{t-1}) } { \sigma_\epsilon^2 + \sigma_\xi^2 }  \\ 
% & = Hg(\x_{t-1}) + \alpha_t (\Z_t - Hg(\x_{t-1})), \\
% G\x^*_t & = Gg(\x_{t-1}),
with $ \lambda_t^* = \f {(\sigma_\xi/\sigma_\epsilon)^2} {(\sigma_\xi/\sigma_\epsilon)^2 + 1} $. % depends on the ratio between state-noise and observation-noise. 
The Hessian matrix of $F(x)$ is
\begin{equation}  \label{Hessian}
\mathrm{Hess}(F)_{i,j} = \left\lbrace
 \begin{matrix}
\sigma_\xi^{-2} + \sigma_\epsilon^{-2},\ &1\leq i=j \leq N_1, \\  
\sigma_\epsilon^{-2},\ \ \ \ &N_1< i=j  \leq N,
\\
0,  & \text{otherwise},
\end{matrix}  \right.
\end{equation}
\blue{In short, the Gaussian distribution $q^{\mathrm{opt}}$ is}
\begin{equation} \label{Implicit_Sampling}
q^{\mathrm{opt}}(\z_t \mid \z_{1:t-1}, \Z_{1:t} ) \sim \mathcal{N} (\x^*_t, \mathrm{Hess}(F)^{-1})
\end{equation}
with $\x^*_t$ in \eqref{MAP_point} and $\mathrm{Hess}(F)$ in \eqref{Hessian}. 

\blue{ In view of feedback control (see e.g.,\cite{auroux2005back,paniconi2003newtonian}), the mean $\x_t^*$ aims to nudge samples to better positions using the observation $z_t$. A  general nudging term is 
\begin{equation} \label{standard nudging equation}
\x_t =g(x_{t-1})+ \lambda_t  \mb M_t  (\Z_t - Hg(x_{t-1})),
\end{equation}
where the real number $ \lambda_t$ represents the strength of nudging, and the nudging matrix $ \mb M_t \in \bb R^{dN \times dN_1}$ provides the direction. 
Thus, $\x_t^*$ can be viewed as a nudging with matrix: $ \mb M_t = [ \mb I_{dN_1\times dN_1};\mb 0_{dN_2\times dN_1}] $ and $\lambda_t  = \lambda_t^*$, optimal in the sense of being the maximizer of the one-step posterior. }

Though optimal for one-step sampling, the above importance density comes with drawbacks: the mean of the unobserved variables, $ G\x^*_t = Gg(\x_{t-1}) $, is simply a projection of the forward equation from the previous state, not updated using information from new observations. % This can also be seen from the nudging matrix in \eqref{standard nudging equation}, in which a block $\mb 0_{dN_2\times dN_1}$ does not provide any updates to the unobserved variables. 
% Empirically, one may seek a nudging matrix that updates the unobserved variables to achieve better performance. 
\blue{Particularly, the next observation $\Z_{t+1}$ is a function of the current unobserved variables $G\x_t$, thus it provides helpful information that we can use to update $G\x_t$. In view of feedback control, this leads to a nudging matrix $\mb M_t$ whose unobserved block containing information from $\Z_{t+1}$. This idea of using future observations has also been explored in auxiliary particle filter (APF) \cite{pitt1999filtering} and lookahead strategies \cite{LCL13}.  Inspired by the APF and the idea of nudging, we propose in the next section an auxiliary sampling strategy with two observations to update the unobserved variables. }

\subsection{Auxiliary sampling with two observations} \label{NIs}
The auxiliary particle filter is an SMC algorithm that makes use of the information from the next observation. To keep the recursive form as in \eqref{eq:rec-Imp}, we need to consider target densities $ p(\z_{1:t} \mid \Z_{1:t+1}) $ instead of $ p(\z_{1:t} \mid \Z_{1:t}) $, and write it recursively as 
\begin{equation*}
\begin{aligned}
p(\x_{1:t} \mid \Z_{1:t+1}) & \propto  \ p(\x_{1:t-1} \mid \Z_{1:t})
\\
& \times \f { p(\x_t\mid\x_{t-1}) p(\Z_t \mid \z_t) p(\Z_{t+1} \mid \z_t)} { p(\Z_t \mid \z_{t-1}) },
\end{aligned}
\end{equation*}

Since the analytical expression of $ p(\Z_{t+1} \mid \z_{t})$ is unknown, we approximate it by $ p(\Z_{t+1} \mid \z_{t}) \approx p(\Z_{t+1} \mid g(\x_{t})) $ and obtain: %Hereafter we abuse the notation by denoting $p(\Z_{t+1} \mid \z_{t}) $ its approximation: 
\begin{equation} \label{eq:rew_res_dis}
\begin{aligned}
\widehat p(\x_{1:t} \mid \Z_{1:t+1}) &\propto  \ \widehat p(\x_{1:t-1}  |\, \Z_{1:t})
\\
& \times \f { p(\x_t|\, \x_{t-1}) p(\Z_t  |\, \z_t) p(\Z_{t+1} |\, g(\z_t))} { p(\Z_t  |\, g(\z_{t-1})) }.
\end{aligned}
\end{equation}

With an importance density $q(\x_t \mid \x_{t-1},\Z_{t:t+1})$ depending on $z_{t+1}$, the recursively updating weight becomes
$ w(\x_{1:t}\mid \Z_{1:t+1}) = w(\x_{1:t-1}\mid \Z_{1:t})  \alpha(\x_{t-1:t},\Z_{t:t:1}) $, where the associated incremental weight is given by:
\begin{equation} \label{eq:weight_use}
\alpha(\x_{t-1:t},\Z_{t:t:1}) = \f { p(\x_t | \x_{t-1}) p(\Z_t | \z_t) p(\Z_{t+1} | g(\z_t))} { p(\Z_t \mid g(\z_{t-1})) q(\x_t \mid \x_{t-1},\Z_{t:t+1})}.
\end{equation}
\vspace*{.1cm}

Next, we construct the importance density $q(\x_t \mid \x_{t-1},\Z_{t:t+1})$ and draw samples from it. 
We start from the negative log function of the posterior distribution  $ p(\Z_{t+1} \mid g(x) p(x \mid \z_{t-1})  p(\Z_t \mid x) $:
\[
\widehat{F}(\x) =  \f {\verti{ \Z_{t+1} - Hg(\x)}^2} {2 \sigma_\xi^2} + \f{\verti{\x - g(\x_{t-1})}^2} {2\sigma_\epsilon^2} + \f {\verti{\Z_t - H\x}^2} {2\sigma_\xi^2}.
\]

Since the state variable is  high-dimensional and its components being indistinguishable agents, it is difficult and computationally costly to find the minimizer of $\widetilde{F}$, who is likely to have multi-modes. This rules out a direct application of implicit sampling. However, by a linear approximation of the nonlinear function $g(\x_t)$, we can directly construct a Gaussian importance density $q(\x_t \mid \x_{t-1},\Z_{t:t+1})$ as the previous section. 
% Because of the nonlinear function $g(\x_t)$, the analytical maximum of $\widehat{F}$ is unavailable. 
We linearize $Hg(\x)$ at $\x_t^*$ since it is the most likely position before the next observation: 
\[
Hg(x) \approx Hg(x_t^*) + \nabla Hg(\x_t^*)^\intercal  (x-x_t^*), 
\]
where $\nabla Hg(\x_t^*) \in \bb R^{dN\times dN_1}$ is the gradient of $Hg$. In practice, when the interaction function $\phi$ is piecewise constant, the approximation of $\grad Hg$ is computed in follows:
\begin{equation}\label{eq:gradHg}
\grad H g(x) \approx \mb R^{IH}(x) + \mb L^{H}(x) \in \bb R^{dN\times dN_1} ,
\end{equation}
where the block matrices $\mb R^{IH}(\cdot)\in \bb R^{dN\times dN_1} $ and $ \mb L^H(\cdot) \in \bb R^{dN_1\times dN_1}$ are composed by submatrices $ \mb R^{II}_{i,j}(\cdot) $ and $\mb L^I_{i,j}(\cdot) \in \bb R^{d\times d} $, respectively:
\[
\mb R^{IH}_{i,j}(x) = \f 1 N \phi(||x^i - H_jx||) I_d,
\]
\[
\mb L^{H}_{i,j}(x) = \left\lbrace 
\begin{aligned}
 &- \f 1 N \mathop \sum_{k=1}^N \phi(||\x^k-\x^i||) I_d ,  \text{ if } 1\leq j = i \leq N_1, \\
 &0 \times I_d, \ \ \  \text{ otherwise }.
\end{aligned} \right.
\]	

Then, $\widehat{F}(\x)$ can be approximated by a quadratic function: 
\[
\begin{aligned}
\widetilde{F}(x) = &  \f {\verti{\Z_{t+1} - Hg(x_t^*) - \nabla Hg(\x_t^*)^\intercal (\x-\x_t^*)}^2} {2\sigma_\xi^2}
\\
& + \f{\verti{x - g(\x_{t-1})}^2} {2\sigma_\epsilon^2} + \f {\verti{\Z_t - Hx}^2} {2\sigma_\xi^2} 
\\
= & \frac{1}{2} y^\intercal Ay - y^\intercal b + C
\end{aligned}
\]
with  $y = \x - \x_t^*$,  $C= \f {  \verti{ \Z_{t+1} - Hg(\x_t^*) }^2 }  {2\sigma_\xi^2} + \f { \verti{g(\x_{t-1}) - \x_t^*   }^2} {2\sigma_\epsilon^2} + \f { \verti{ \Z_t - H\x_t^* }^2 }  {2\sigma_\xi^2}$, and
\begin{equation} \label{eq:yAb}
\left\lbrace\begin{aligned}
A = &  \f {\nabla Hg(\x_t^*) \nabla Hg(\x_t^*)^\intercal } {\sigma_\xi^2} + \f {I_N} {\sigma_\epsilon^2} + \f {H^\intercal H} {\sigma_\xi^2},
\\
b = &   \f {\nabla Hg(\x_t^*)^\intercal  \left[ \Z_{t+1} - Hg(\x_t^*) \right] } {\sigma_\xi^2} +  \f { g(\x_{t-1}) - \x_t^* } {\sigma_\epsilon^2}
\\
& + \f {H^\intercal \left[\Z_t - Hx_t^*\right]} {\sigma_\xi^2}. 
\end{aligned}\right. 
\end{equation} 
Then, $\widetilde{F}$ has a minimizer $\mu(\z_{t-1}, \Z_t, \Z_{t+1})$ given by: %(when not casing confusion, simply denoted as $\mu$)
\begin{equation} \label{eq:mu}
\mu(\z_{t-1}, \Z_t, \Z_{t+1}) = \x_t^* +A^{-1} b
\end{equation} 
and its Hessian is $A$. This suggests the following importance density $\impd(\z_t\mid \z_{t-1}, \Z_t,\Z_{t+1})$:
\begin{equation} \label{importanceDensity}
\impd(\z_t\mid \z_{t-1}, \Z_t, \Z_{t+1}) \sim \mathcal{N} (\mu(\z_{t-1}, \Z_t,\Z_{t+1}) , A^{-1} ),
\end{equation}
where $\mu(\z_{t-1}, \Z_t,\Z_{t+1})$ is defined by \eqref{eq:mu} and $A= A(\z_{t-1}, \Z_t) $ is defined by \eqref{eq:yAb}.

We summarize the above in the following algorithm:
\begin{algorithm}[!htbp]
	\caption{Auxiliary implicit sampling} \label{alg:AIS}
	At time $ t \leq T-1 $, for $ s = 1,2,...,S $, do:
	\begin{itemize}
		\item Evaluate $x_t^* = x^*(x_{t-1}^{(s)},\Z_t )$ as in \eqref{MAP_point}, and then compute $A(\z_{t-1}^{(s)}, \Z_t)=A$ as in  \eqref{eq:yAb}  and $\mu(\z_{t-1}^{(s)}, \Z_t,\Z_{t+1}) $ as in \eqref{eq:mu}.
		\item Draw a sample $\z_t^{(s)}$ from a normal distribution with mean $ \mu(\z_{t-1}^{(s)}, \Z_t,\Z_{t+1},) $ and covariance $ A(\z_{t-1}^{(s)}, \Z_t)^{-1} $; evaluate weights $ \widehat {w}_t^{(s)} = w_{t-1}^{(s)} \cdot \alpha_t^{(s)} $ as in \eqref{eq:weight_use}.
		\item Resample to obtain equally-weighted samples (if a criterion is met). 
	\end{itemize}
	At time $t = T$, for $ s = 1,2,...,S $, do Implicit sampling:
	\begin{itemize}
		\item Evaluate $x_t^* = x^*(x_{t-1}^{(s)},\Z_t )$ as in \eqref{MAP_point}.
		\item Draw a sample $\x_t^{(s)}$ from $\mathcal{N} (\x^*_t, \text{Hess}(F)^{-1})$ as in \eqref{Implicit_Sampling}; evaluate weights $ \widehat {w}_t^{(s)} $ by \eqref{eq:weight}.
	\end{itemize}
\end{algorithm}

\subsection{MCMC-move} \label{Mmove}
To further reduce the inevitable degeneracy of SMC algorithms, we introduce an MCMC-move step \cite{berzuini1997dynamic}. 
We consider two types of moves: a directional move aiming for agent-wise position improvement, and a local trajectory move aiming to replace a low weight short-trajectory by a higher weighted one

The \textbf{directional move} randomly selects $m$ of the unobserved agents for each sample,  and resample each of them using a Metropolis-Hastings step as follows.  For each selected agent $k$, first draw a sample from $\mathcal{N}(x^k_*, \Sigma_*^k)$, where $x^k_*$ is a minimizer of the function 
\[ 
\widetilde g\,(\x_t^k | \Z_{t+1}, \x_t^{1:k-1,k+1:N}) =  \|\Z_{t+1}- Hg(\x_t)\|^2,
\]
and $ \Sigma_*^k$ is the Hessian of the function $\widetilde g$ at $x^k_*$, that is, 
\begin{equation}\label{eq:g_move}
\x^k_* = \argmin{x\in\R^d} \widetilde g(x); \quad \Sigma^k_* =  \mathrm{Hess}\, \widetilde g\,(\x^k_*). 
\end{equation}
Then,  accept the sample if it leads to a higher likelihood for observation $\Z_t$. We set the number $m$ to be $ \lfloor \beta N_2 \rfloor $ (with $\beta = 0.2$). \blue{In practice, the optimization can be relaxed to a few iterations of gradient descent search, since the goal of our directional move is only to improve the position of partial agents. Multiple Try Metropolis (MTM) methods \cite{liu2000multiple,martino2013flexibility,pandolfi2010generalization} are good alternatives, particularly when the gradient is not available for the optimization.  }

We summarize the directional move in Algorithm \ref{alg:DirectionMove}. 

\begin{algorithm}[!htbp] 
	\caption{Directional move} \label{alg:DirectionMove}
	At time $ t \in \text{checking-time} \subset \{ 1,...,T \} $, for each sample, do:
	\begin{itemize}
		\item Randomly select $m=  \lfloor 0.2 N_2 \rfloor $ of the unobserved agents.
		\item Move the selected agents: for $ k = 1,...,N_2 $,  if the agent is among those selected, sample $ \widetilde{\z}_{t}^{k} \sim \mathcal{N}(x^k_*, \Sigma_*^k)$,  where $ x^k_*$ and $\Sigma_*^k$ are defined in \eqref{eq:g_move}; 			else, set $ \widetilde{\z}_{t}^{k} = \z_{t}^{k}$. 		
		\item Accept the move and set $ \z_{1:t}' = [\z_{1:t-1},\widetilde \z_t] $ if $ u \sim \mathcal{U}_{[0,1]} \leq \min \left\lbrace 1, \f {p(\Z_t \mid \widetilde \z_t)} {p(\Z_t \mid \z_t)} \right\rbrace $; otherwise, reject the move and set $ \z_{1:t}' = \z_{1:t} $
	\end{itemize}
\end{algorithm}

The \textbf{local trajectory move} randomly selects low-weighted samples and replaces their local trajectories by those with a higher probability. More precisely, at a prescribed time, a sample with index $s$ is selected with probability $ \max \left\lbrace 0, 1- \ff {w_t^{(s)}} {c_t} \right\rbrace $, where $c_t$ is the value of the lowest quartile of the weights $\{w_t^{(s)}\}_{s=1}^S$. Intuitively speaking, all samples with weight higher than $c_t$ will be kept and a sample with weight less than $c_t$ will be selected randomly, according to a probability that increases when its weight decreases.
Once selected, its local trajectory $x^{(s)}_{t-T_0:t}$ is moved to $ \widetilde{\x}_{t-T_0:t}^{(s)} $ as in Algorithm \ref{algorthm: Local-trajectory move}.

\begin{algorithm} [!htbp] 
	\caption{Local-trajectory move}
	\label{algorthm: Local-trajectory move}
	At time $ t \in \text{checking-time} \subset \{ 1,...,T \} $, with samples $\{ \x_{1:t}^{(s)}, w_t^{(s)}\}_{s=1}^S$ and the weights $\{w_{t-T_0}^{(s)}\}_{s=1}^S$, do:
	\begin{itemize}
		\item Select low-weighted samples: for $ s \in \{ 1,...,S \} $, set an indicator $ \Theta_t^{(s)} =1 $ with probability  $ \max \left\lbrace 0, 1- \ff {w_t^{(s)}} {c_t} \right\rbrace $, where $c_t$ is the value of the lowest quartile of the weights $\{w_t^{(s)}\}_{s=1}^S$;
		\item Move the low-weighted samples: for $ s \in \{ 1,...,S \} $, if $ \Theta_t^{(s)} = 1$, replace the local trajectory  ${\x}_{t-T_0:t}^{(s)}$ by as follows: 
		\begin{itemize}
			\item Draw a sample $ \widetilde{\x}_{t-T_0}^{(s)} $ from the samples $\{ \x_{t-T_0}^{(s)}, w_{t-T_0}^{(s)}\}_{s=1}^S$; 
			\item Implement a directional move for $ \widetilde{\x}_{t-T_0}^{(s)} $ as in Algorithm \ref{alg:DirectionMove}, in which, draw new positions for $m$ of the unobserved agents from the initial distribution, instead of drawing samples from $\mathcal N(\x_*^k, \Sigma_*^k))$; 
			% by marginal distribution of initial distribution $\mu$.
			\item	Draw $ \widetilde{\x}_{t-T_0:t}^{(s)} $ by a one-sample SMC algorithm with importance density function in \eqref{importanceDensity} from $t-T_0$ to $t$ with initial value $  \widetilde{\x}_{t-T_0}^{(s)}  $ ;
			\item Accept the move and set $\z_{t-T_0:t}^{(s)} = \widetilde{\x}_{t-T_0:t}^{(s)}  $, $ w_t^{(s)} = c_t $ if $ u \sim \mathcal{U}_{[0,1]} \leq \min \left\lbrace 1, \f {p(\Z_t \mid \widetilde \z_t)} {p(\Z_t \mid \z_t)} \right\rbrace $; otherwise, reject the move and keep them as original. 	
		\end{itemize}
	\end{itemize}
\end{algorithm}

\subsection{Rejection of non-physical samples} \label{Imove}
To avoid non-physical samples, we introduce an \textbf{information move} step, rejecting non-physical samples. We say a sample is non-physical if it violates the basic properties of the opinion dynamics. For example, recall the following contraction of radius property of opinion dynamics  \cite[Proposition 2.1]{motsch2014heterophilious}: for any constant $c \in \bb R^d$, we have:
\[
\max_i \|\x_{t}^i - c\| \leq \max_i \|\x_{t'}^i - c\| , \ \ \forall t \geq t'.
\]
This property requires information of all agents, and it can not be directly applied to our partial observations. Since the mean position of all agents does not change in time, we say a sample at time $t$ is non-physical if
\begin{equation}\label{eq:non_physical}
\max_i \|\x_{t}^i - \overline{\x}_{t-t_0}\| > \max_i \|\x_{t-t_0}^i - \overline{\x}_{t-t_0}\| + \alpha , \ \ \forall t \geq t'.
\end{equation}
where $t_0$ and $\alpha>0$ are constants (in practice, $t_0 = 10$ and $\alpha = 0.3 \times \text{supp}(\phi)$), representing the time length we back-track for checking and the tolerance for extending the maximal distance, respectively.

We also reject agents that do not interact with any of the observed agents. Such agents may be connected to the observed agents through other unobserved agents (recall that $\x_t^i$ and $\x_t^j$ are connected if there exists a path of agents $\{ \x_t^{i_k} \}_{k=0}^K$ with $i_0=i$ and $i_K=j$ such that $\phi(\| \x_t^{i_k} - \x_t^{i_{k+1}} \|)> 0 $), but their positions are difficult to estimate from the limited information. It is of interest to replace them by agents that interact with the observations: if they evolve to disconnect from the observed agents, their clustering can not be estimated from the observations, thus we can view them as ``non-physical'' (or with little information); otherwise, they will move toward the center of the cluster, and the replacement will accelerate their move.

In addition, to avoid over-correction and to maintain computational efficiency, we apply the rejection-moves at a pre-specified time steps that performed progressively less frequently as observation increases. 

The information move algorithm is summarized in Algorithm \ref{alg:InfoMove}:
\begin{algorithm}[!htbp]
	\caption{Information move}\label{alg:InfoMove}
	For $ t\in \text{checking-time} \subset \{ 1,...,T \} $ and for each sample, do:
	\begin{itemize}
		\item For $j=1,...,N_2$, check if $\z_{t}$ is non-physical as in \eqref{eq:non_physical} or if it has agents disconnected from the observed agents. If yes, draw a sample $\widetilde \z_{t} $ from the importance density in \eqref{importanceDensity} and repeat until the sample is physical and connected with the observed agents.% ; else set $ G_j\z_{t}' = G_j\z_{t} $.
		\item If $ u \sim \mathcal{U}_{[0,1]} \leq \min \left\lbrace 1, \f {p(\Z_t \mid \widetilde \z_t)} {p(\Z_t \mid \z_t)} \right\rbrace $, set $ \z_{t} = \widetilde \z_t $; else, repeat from the previous step until accepted.
	\end{itemize}
\end{algorithm}

\subsection{Summary}
We combine all the above sampling techniques into Algorithm \ref{alg:ourSMC}, which we refer it as auxiliary implicit sampling (AIS). 
\begin{algorithm}[!htbp]
	\caption{Auxiliary implicit sampling with MCMC moves (AIS)}\label{alg:ourSMC}
	\textbf{At} time $t = 1$, initialization: draw uniform-weighted samples $\{ \z_1^{(s)}, w_0^{(s)}\}$ from $\mu(\z_1)$. 
	
	\textbf{For} time $t \geq 2$, do:
	\begin{enumerate}
		\item [(a)] Draw weighted sample $\{{\z}_{1:t}^{(s)},  w_t^{(s)}\}$ by the Auxiliary implicit Sampling Algorithm \ref{alg:AIS}. 
		\item [(c)] Improve the samples by two MCMC moves: the Directional Move Alorithm \ref{alg:DirectionMove} and the Local Trajectory Move Algorithm \ref{algorthm: Local-trajectory move} when resampling occurs. 
		\item [(d)] Reject non-physical samples by the Information Move Algorithm \ref{alg:InfoMove} when resampling occurs. 
	\end{enumerate}
\end{algorithm}

%%%%%%%%%%%%%%%%%=====================
%%%%%%%%%%%%%%%%%=====================
%%%%%%%%%%%%%%%%%=====================
\section{Numerical experiments} \label{Simulation}
In this section, we predict the clustering of the opinion dynamics using partial observations, following the Bayesian approach discussed in Section \ref{sec:Bayes}, using the auxiliary implicit sampling (AIS) algorithm introduced in Section \ref{sec:Method}.  We first describe the settings of the model and the sampling method in Section \ref{sec:settings}.  Then, we present results on state estimation in Section \ref{sec:stateEst}. We report the prediction of clustering in a typical simulation in Section \ref{Cp1} and in many simulations in Section \ref{Cp100}.

\subsection{Numerical settings} \label{sec:settings}
We consider the opinion dynamics \eqref{opinion dynamic} with $N=60$ agents, 
\[
\x_{t+1}^i - \x_{t}^i = \f 1 N \mathop \sum_{j=1}^N \phi(\|\x_{t}^j-\x_{t}^i\|)(\x_{t}^j-\x_{t}^i) \Delta t
\]
where $\x_{t}^i \in \R^d$ with $d=2$ represents the opinion of the agent $i$ at discrete times indexed by $t$. This system is an Euler approximation of the corresponding differential equations with time step size $ \Delta t = 0.05$.

\begin{figure}[!t] 
	\centering
	\begin{subfigure}[b]{1\linewidth}	\centering
		\includegraphics[width=0.48\linewidth]{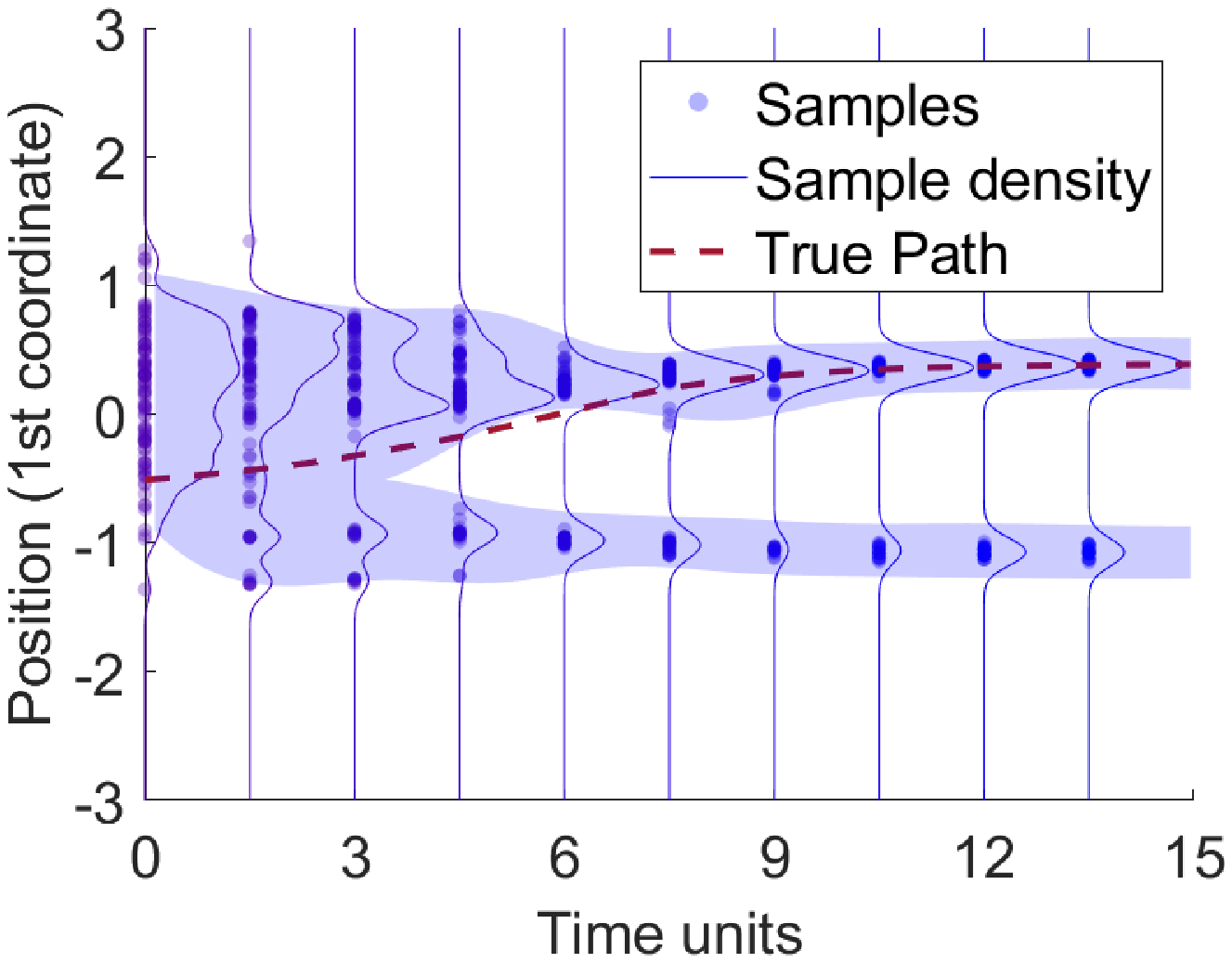}
		\hfil
		\includegraphics[width=0.48\linewidth]{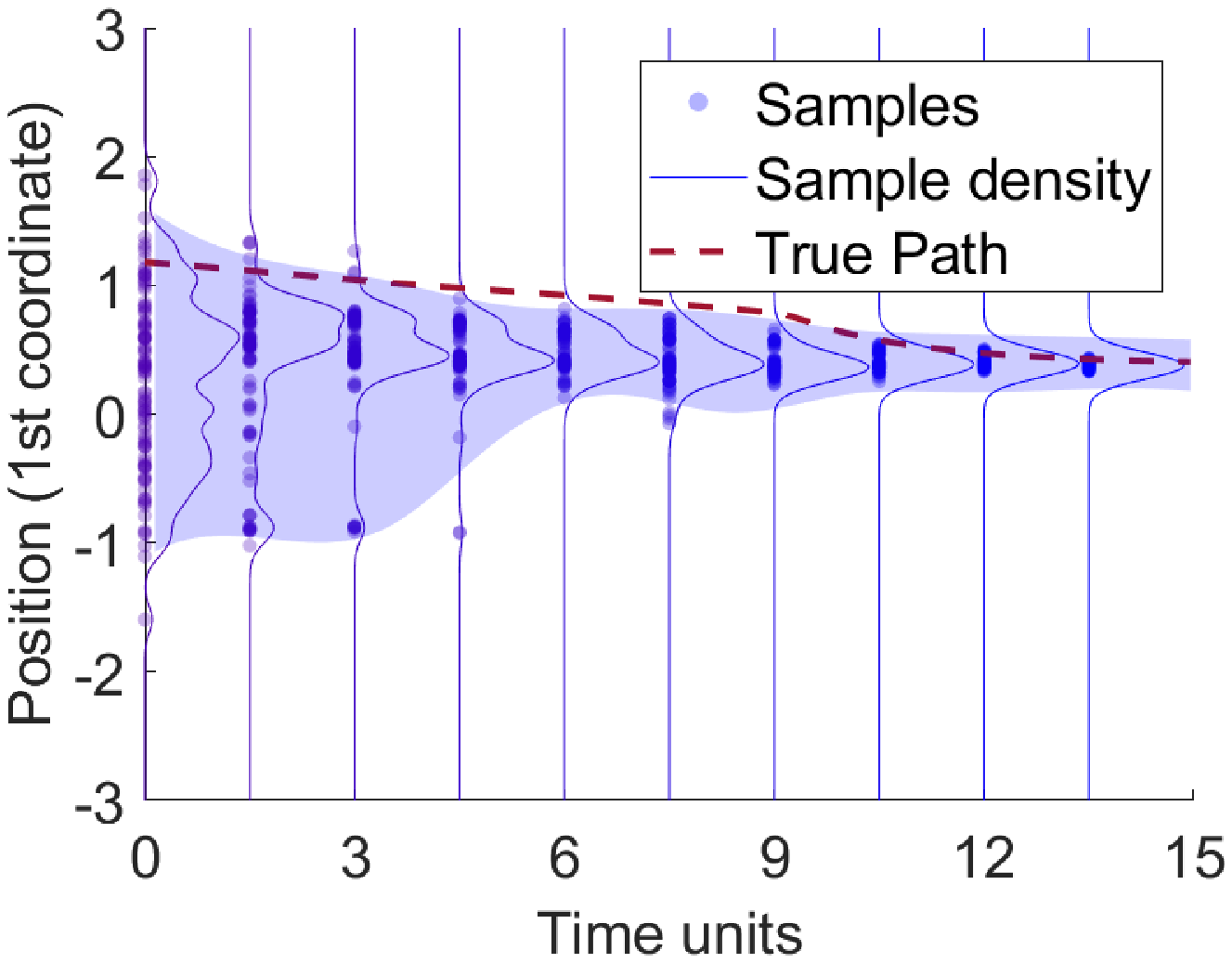}
		\hfil
		\caption{The 1st coordinate of two unobserved agents}
		\label{SeqEvlSingle2}
	\end{subfigure}\hspace{-0.09in}
	\quad
	\begin{subfigure}[b]{0.7\linewidth}
		\includegraphics[width=1\linewidth]{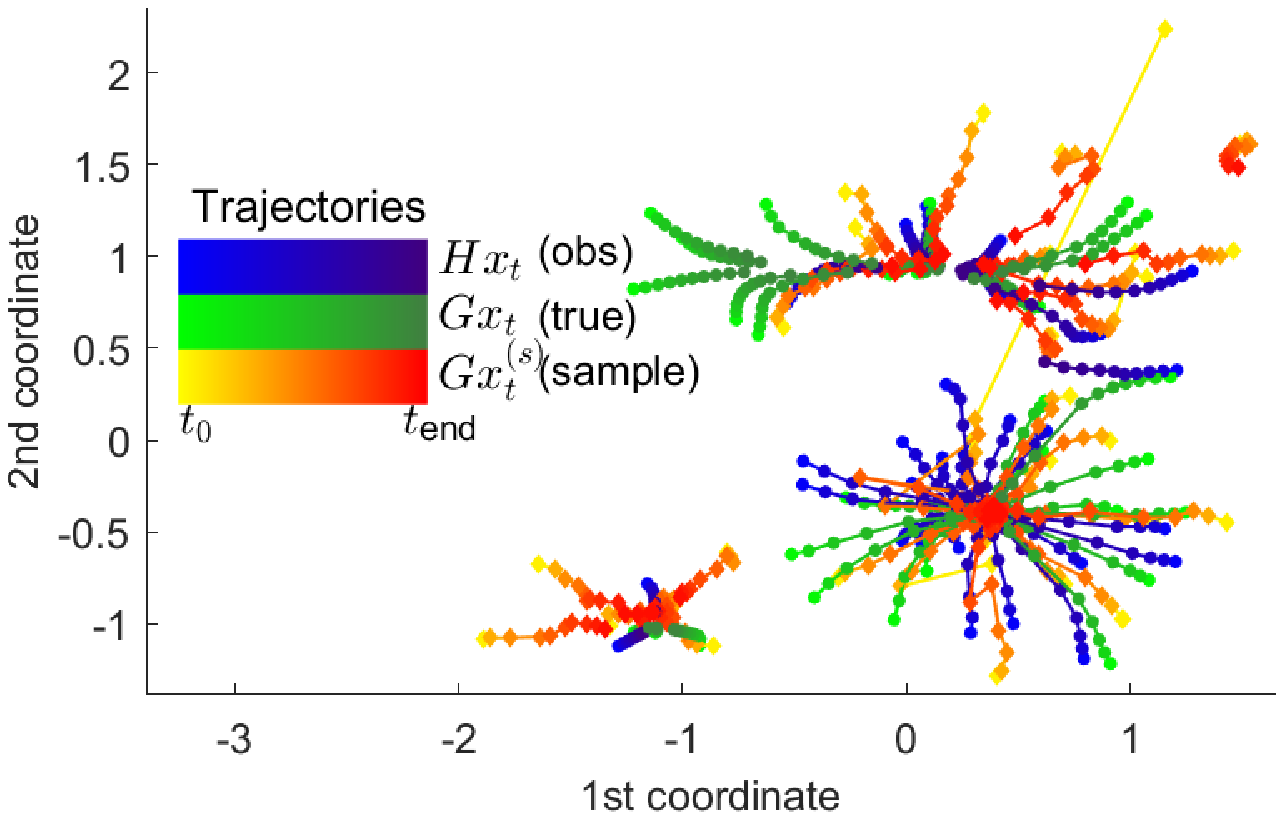}
		\caption{Trajectories of all agents: truth and a sample} 
		\label{SeqEvlSingleall1}
	\end{subfigure}
	\caption{State estimation (noiseless observations): Estimation of the trajectory of agents for system without noise, observing $N_1=30$ of the $N=60$ agents. (\ref{SeqEvlSingle2}) shows the paths of the first coordinate of two unobserved agents for all the $S=100 $ samples (blue dots). At each time, the blue curve is the smoothed empirical marginal posterior density from the samples (Sample density), and the blue shaded area is the 95\% credible interval. For each agent, samples become concentrated around the truth (the red dash line) as time increases, with the marginal posterior peaks near the truth. The sample density may have multiple modes and the true value is in the 95\% credible interval for most of the times. (\ref{SeqEvlSingleall1}) shows the trajectories of all agents, where the blue and green dots are the observed and unobserved truth, and the red diamonds are the unobserved agents in a sample; all with color changing from light to dark as time increases. The estimated trajectories of unobserved agents by the sample can be far away from the truth, particularly at the initial time, but the clustering of the sample is close to the truth. }
	\label{Noiseless,type1}
\end{figure}

\blue{Since we are interested in the cases when the system formulates multiple clusters instead of a consensus, }we consider a communication function $\phi$ that is piecewise-constant:
\[
\phi(r) =  \left\lbrace\begin{aligned}
&1 , \ \ \ \ \ \ \ &r \in [0,\sqrt{2}/2),
\\
&0.1 , \ \ \ &r \in [\sqrt{2}/2,1),
\\
&0 , \ \ \ & r \in [1,\infty).
\end{aligned}\right.
\]
\blue{This local communication function represents a stronger interaction between alike-opinions than different-opinions, and it is more likely to lead to multiple clusters than heterophilious interactions \cite{motsch2014heterophilious}.}
We obtain multiple clusters by selecting the initial conditions as follows: we randomly draw initial condition for each agent from $ \mu \sim \mathrm{Unif}([-4,4]^d)$, %in other words, the initial distribution of $(x_0^1,\ldots, \x_0^N)$ is $\mu^{\otimes N}$ on $\R^{dN}$,
 and reject those leading to consensus. We will call the empirical distribution of these selected initial conditions as initial distribution of the opinion dynamics. This initial distribution injects randomness into the dynamics.

Our goal is to predict the clustering of the system, particularly the sizes and the locations of the largest clusters, supposing that we only observe the trajectories of $N_1$ of the $N$ agents for a relatively short time, far before the clusters are formulated. In particular, we assume that we observe the system for only $n=300$ time steps, when the observations can not tell if the clusters have formulated. The clustering usually takes more than 30 time units, or equivalently 600 time steps. (For instance, the system described by Figure \ref{Prediction for clusters noiseless} clustered at about 1500 time steps.)  As discussed in Section \ref{sec:Bayes}, a Bayesian approach provides a probabilistic framework for state estimation and cluster prediction, with uncertainties quantified by the posterior. We sample the posterior by the auxiliary implicit sampling (AIS) algorithm in Algorithm \ref{alg:ourSMC} with ensemble size $S=100$. For the sake of the AIS, we rewrite the system in the form of a state-space model:
\[
\left\lbrace \begin{aligned}
\z_{t+1} & = g(\z_t) + \epsilon_t, \ \ \ \z_1^i \sim \mu , \forall i,
\\
\Z_{t} & = H\z_t + \xi_t,
\end{aligned} \right.
\]
where \red{$ \epsilon_t \sim \mathcal{N}(0, \sigma_\epsilon^2 I_{dN})$ and $ \xi_t \sim \mathcal{N}(0, \sigma_\xi^2 I_{dN_1}) $.} 

We consider systems with both noiseless and noisy observations. To avoid degenerate distributions, we set an artificial noise for the deterministic state model. For the case of noiseless observations, \red{we set $ \sigma_\epsilon = 0.01 $ and $ \sigma_\xi = 0.005$, so that the artificial noise is relatively small with respect to the signal. For noisy observations with $\sigma_\xi =0.01$ (which represents a signal-to-noise ratio about 2\%), we set $\sigma_\epsilon = 0.05$. 
In both cases, we have $\sigma_\xi/\sigma_\epsilon <1$ so that the important densities trust the state model more than the observations while keeping relatively large variances, see \eqref {MAP_point}--\eqref{Implicit_Sampling} and \eqref{eq:yAb}--\eqref{importanceDensity}.   }

\begin{figure}[!t]
	\centering
	\begin{subfigure}[b]{1\linewidth}	\centering
		\includegraphics[width=0.48\linewidth]{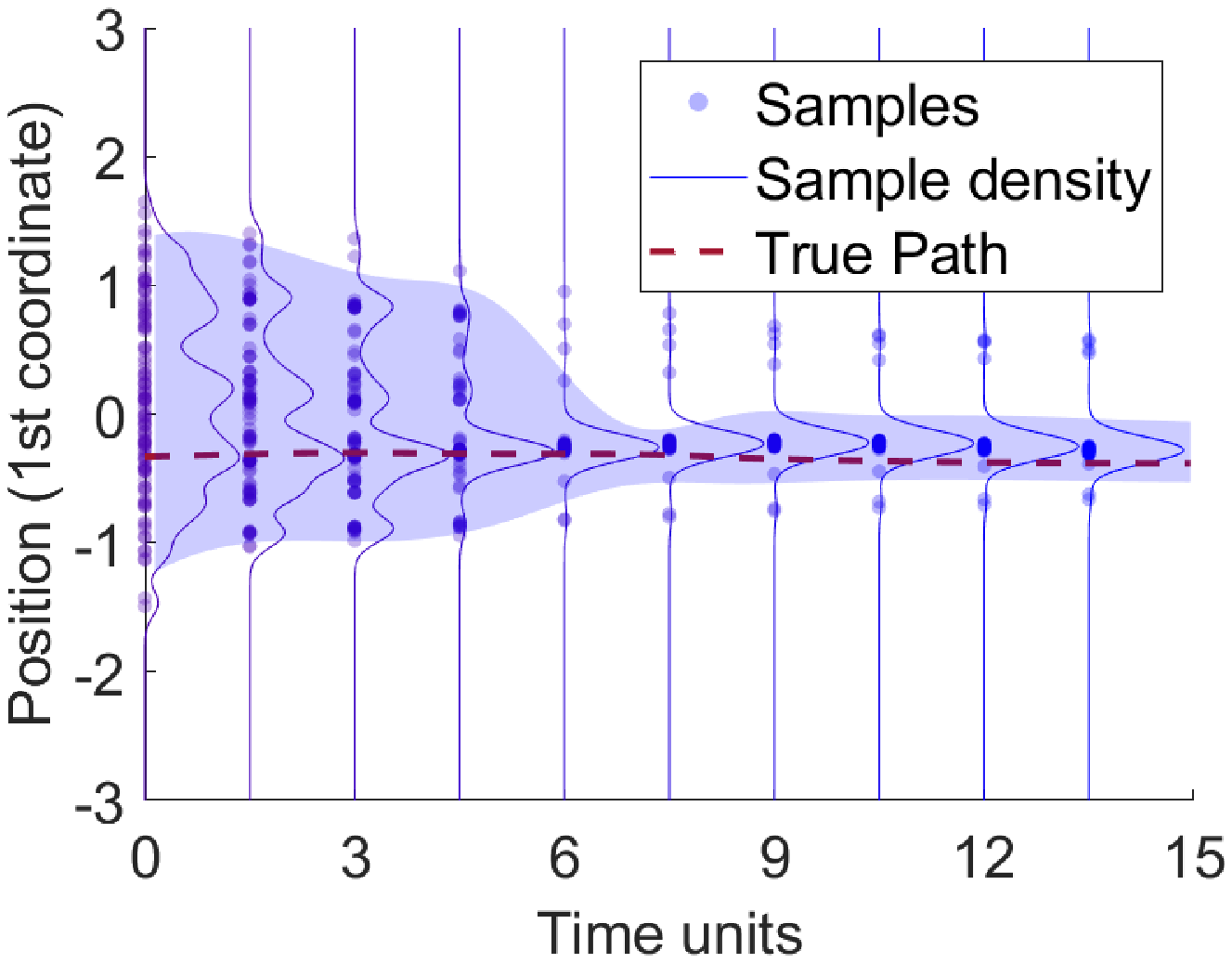}
        \hfil
		\includegraphics[width=0.48\linewidth]{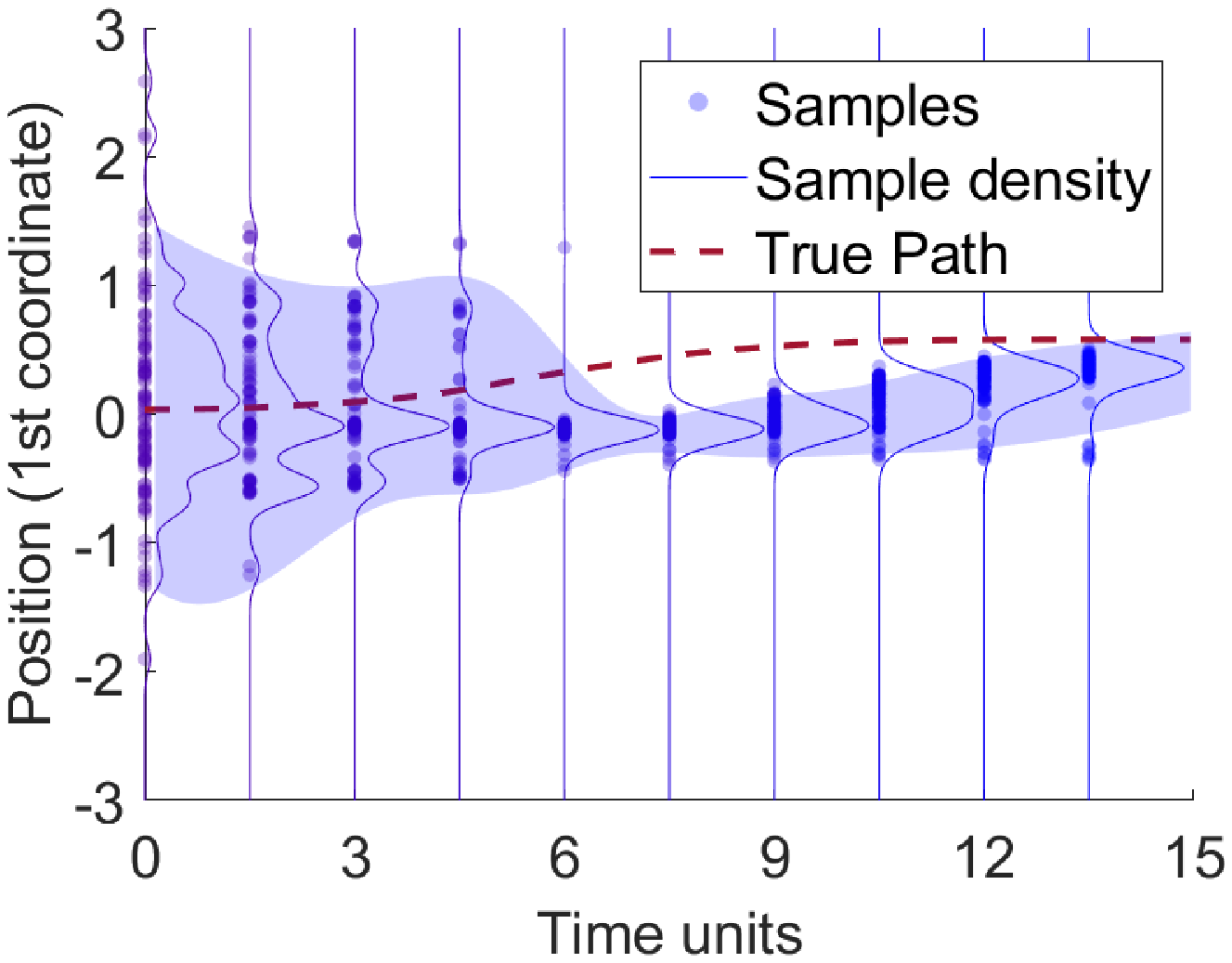}
		\hfil
		\caption{The 1st coordinate of two unobserved agents}
		\label{SeqEvlSingle4}
	\end{subfigure}
	\begin{subfigure}[b]{0.7\linewidth}
		\includegraphics[width=1\linewidth]{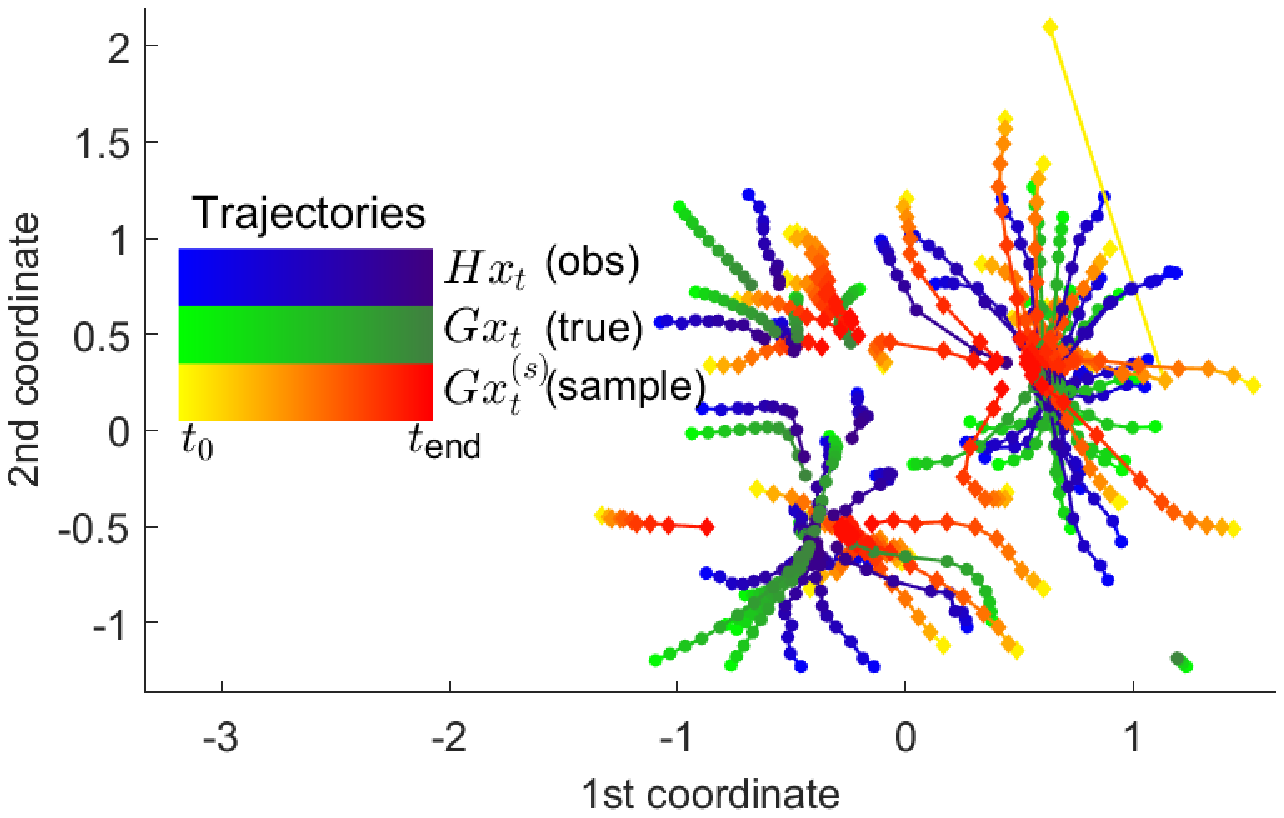}
		\caption{Trajectories of all agents: truth and a sample} 
		\label{SeqEvlSingleall2}
	\end{subfigure}
	\caption{State estimation (noisy observations): Estimation of the trajectory of agents when observing $N_1=30$ of the $N=60$ agents with additive Gaussian noise. In (\ref{SeqEvlSingle4}), the samples (blue dots) of the 1st coordinate become concentrated around the truth (the red dash line) as time increases, with the multi-mode sample density (blue line) peaks near the truth. The blue shaded area is the 95\% credible interval, covering the true value for most of the times.  Due to the random observation noises, some of the true values fall out of the credible interval in the right figure. 
	(\ref{SeqEvlSingleall2}) shows the trajectories of all agents, where the blue and green dots are the observed and unobserved truth, and the red diamonds are the unobserved agents estimated by a sample; all with color changing from light to dark as time increases. The estimated trajectories of unobserved agents by the sample can be far away from the truth, particularly at the initial time, but the clustering of the sample is close to the truth.  }
	\label{Noisy,type1}
\end{figure}

%%%%%%%%%%%%%%%%%================ 
\subsection{State estimation}\label{sec:stateEst}
As a Bayesian approach, our goal of state estimation is to represent the posterior of the states, which is approximated by the empirical measure of the samples in our sequential Monte Carlo algorithm. We demonstrate the state estimation by the marginal posteriors of the trajectories of the first coordinate of two unobserved agents. We also show the trajectories of all agents, comparing the estimated path of unobserved agents in a sample with the truth.

\paragraph{Noiseless observations} Consider first the case when the system is deterministic and half of the $N=60$ agents are observed without noise. Figure \ref{SeqEvlSingle2} shows the trajectories of all the $S=100$ samples for the first coordinate of two unobserved agents, along with the smoothed sample density at each time, representing the marginal posterior. For each agent, samples become concentrated around the truth (the red dash line) as time increases, with the marginal posterior peaks near the truth. Such a concentration of the sample agrees with the intuition that the uncertainty in the posterior of the states decreases when more observations are available, since the system is deterministic and the randomness comes only from the initial condition. The marginal posterior has multiple modes, reflecting the symmetry between agents and the non-identifiability of the states as discussed in Section \ref{Observability}. 

Figure \ref{SeqEvlSingleall1} shows the trajectories of all agents, comparing the estimated paths of unobserved agents estimated by a sample with the truth. The estimated trajectories by the sample can be far away from the truth, particularly at the initial time, but the clustering of the sample is close to the truth.

\paragraph{Noisy observations} We also consider the case when half of the $N=60$ agents are observed with additive Gaussian noise $ \mathcal{N}(0, \sigma_\xi^2 I_{dN_1}) $. Similarly, Figure \ref{SeqEvlSingle4} shows the trajectories of all the $S=100$ samples for the first coordinate of two unobserved agents, along with the smoothed empirical sample density at each time, representing the marginal posterior. For each agent, the sample density is more wide-spread and has more modes than those in Figure \ref{Noiseless,type1} for the deterministic system, indicating more uncertainty due to the noises in the system and in the observation. But all samples become concentrated around the truth eventually as time increases, with the marginal posterior peaks near the truth.  

Figure \ref{SeqEvlSingleall2} shows the trajectories of all agents, comparing the estimated paths of unobserved agents estimated by a sample with the truth. Again, the estimated trajectories by the sample can be far away from the truth, particularly at the initial time, but the clustering of the sample is close to the truth. 

%\bigskip 
In summary, in both noiseless and noisy observations, the marginal posteriors of the state can be multi-mode, presenting a large uncertainty; the trajectories of the samples can be far from the truth,  but the clustering pattern of the samples is close to the truth.

\paragraph{Efficiency of the AIS algorithm} We assess the efficiency of the AIS algorithm by the effective sample sizes and the frequency of resampling. A high ratio of ESS suggests that the samples are close to uniformly weighted so that the importance density is close to the target density. A low frequency of resampling indicates a slow pace of degeneracy in the samples. Together they indicate the efficiency of the SMC algorithm. 

\begin{figure}[!t]
	\centering
	\includegraphics[width=0.48\linewidth]{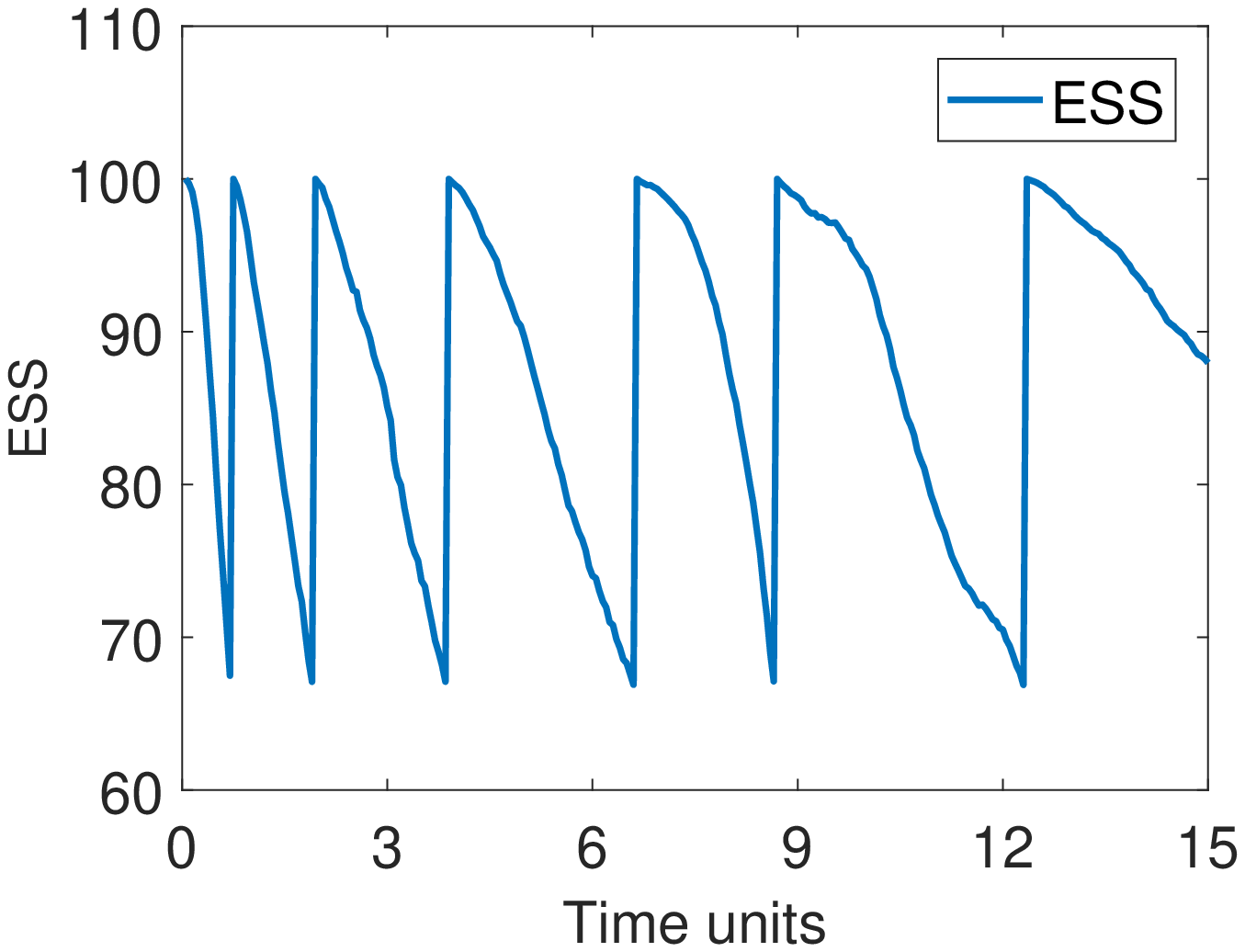}
	\hfil
	\includegraphics[width=0.48\linewidth]{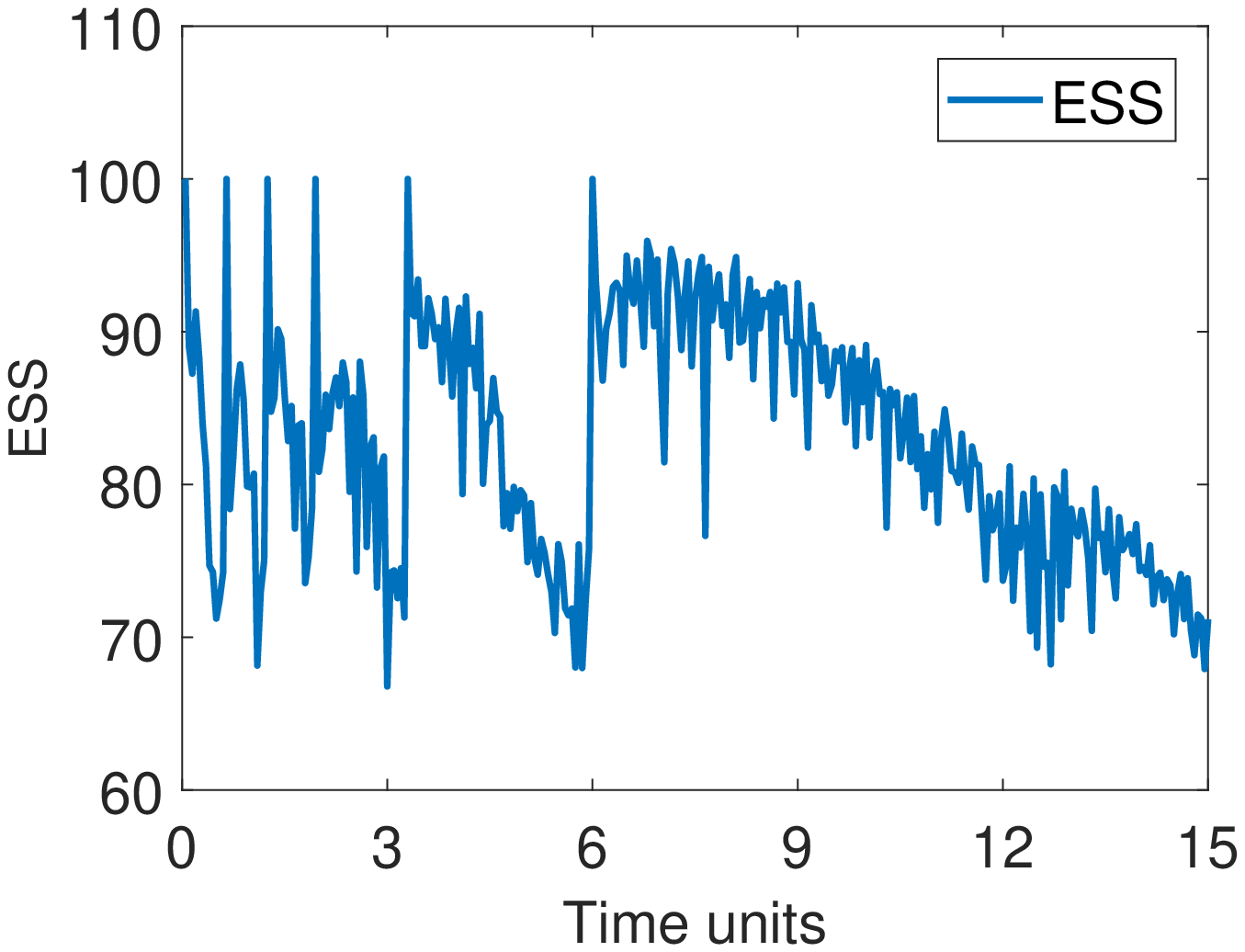}
	\hfil
	\caption{ Effective Sample Sizes (ESS) of AIS in the typical simulations: noiseless observation (left) and noisy observation (right). The ESS drops down to the resampling threshold (set to be 67) for only a few times along the trajectory of 300 time steps, suggesting the high efficiency of the AIS. }
	\label{ESS_figs}
\end{figure}

Figure \ref{ESS_figs} presents the ESS's of the above typical simulations with noiseless and noisy observations. The ESS drops slowly to the resampling threshold (set to be 67) in both cases. The resampling occurs only 6 times along the trajectory of 300 time steps in the case of noiseless observation, and this number drops to 3 in the case of noisy observation. Thus, the AIS is a highly efficient SMC algorithm. 

Also, the AIS is computationally efficient. Since its importance densities are derived analytically, the core AIS in Algorithm \ref{alg:AIS} does not invite any additional computational cost beyond the necessary forward solutions of the state model to generate samples. Extra costs occur when we add the MCMC-move and information-move in Algorithm \ref{alg:DirectionMove}-\ref{alg:InfoMove}. But these costs can be controlled and we only apply them when resampling occurs.

%%%%%%%%%%%%%%%%==============
%%%%%%%%%%%%%%%%==============
%%%%%%%%%%%%%%%%==============
\subsection{Clustering prediction: a typical simulation} \label{Cp1}
In this and the next section, we consider the prediction of clusters from partial observations. We exhibit the cluster prediction of a typical simulation in this section, and we report the performance in many simulations in the next section.

Recall that in a typical clustering prediction, we characterize the clustering by the posteriors of the sizes and centers of the clusters, particularly the leading clusters. We compare our AIS algorithm with two other SMC algorithms: SIR and implicit sampling (denoted by IS). Since the degeneracy of samples in SIR  is too severe for any meaningful prediction, we reduce the degeneracy by inflating its weights to keep more samples in each step. For the prediction of the sizes, we also compare AIS with the predictions simply based on connected agents from observation $\x_T$ only, since in practice one may treat the observation as a random sample of the population.

\begin{figure}[!t]
	\centering
	\begin{subfigure}[b]{1\linewidth}	\centering
		\includegraphics[width=0.48\linewidth]{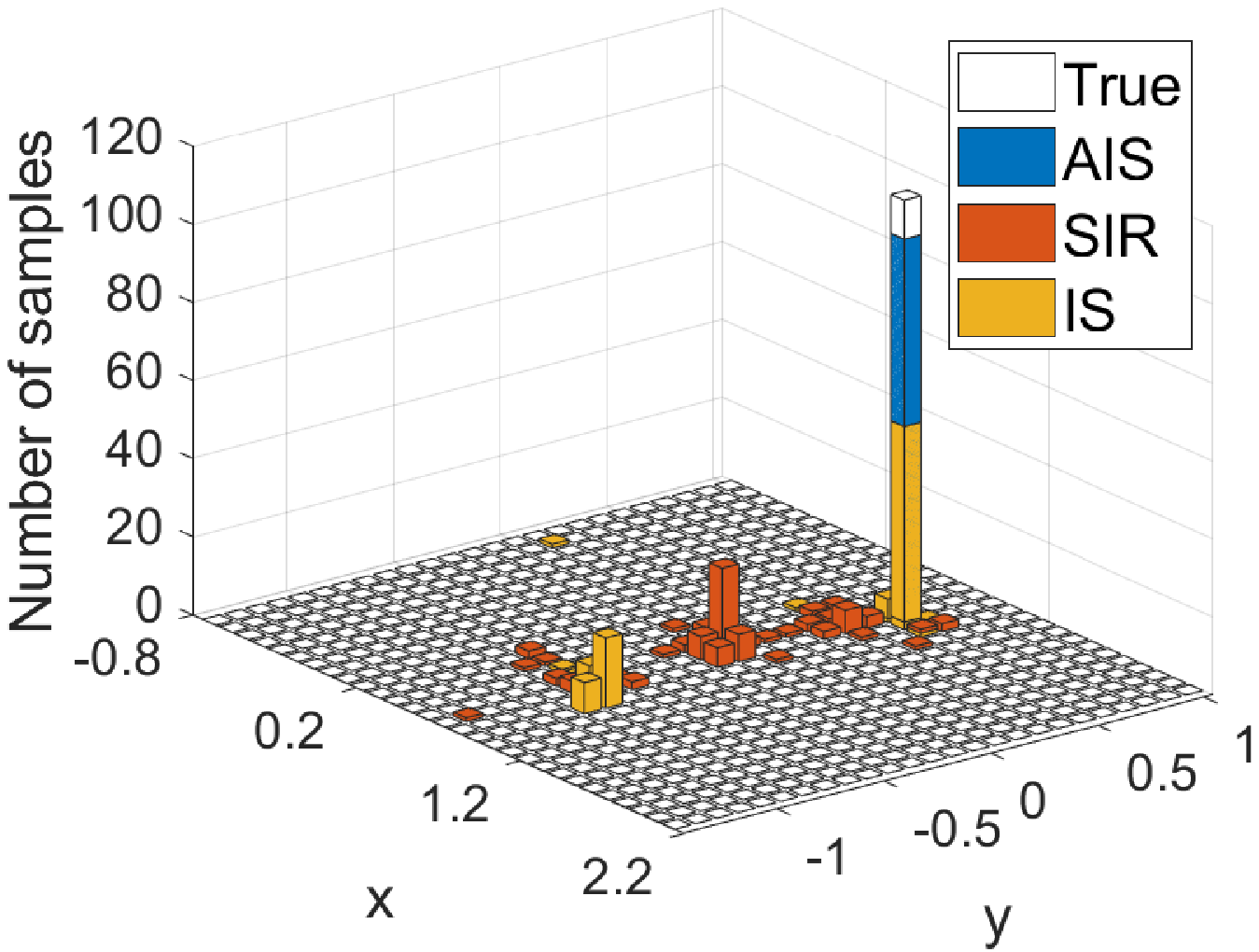}
		\hfil
		\includegraphics[width=0.48\linewidth]{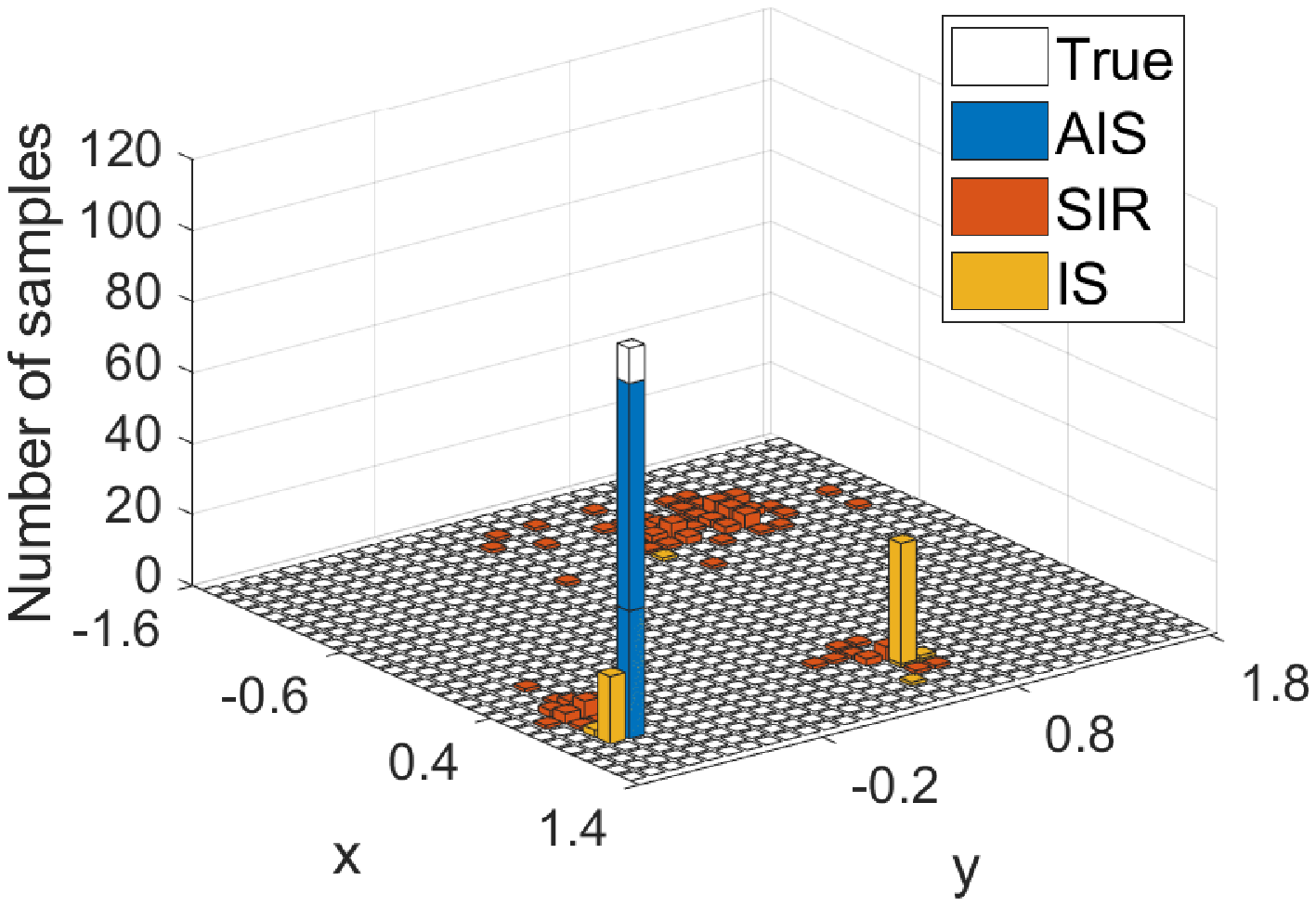}
		\hfil
		\caption{Posterior of cluster center: the  largest (left) and 2nd largest (right) cluster}
		\label{ComparePolar0}
	\end{subfigure} 
	\quad
	\vspace{.2cm}
	\begin{subfigure}[b]{1\linewidth}	\centering
		\includegraphics[width=0.485\linewidth]{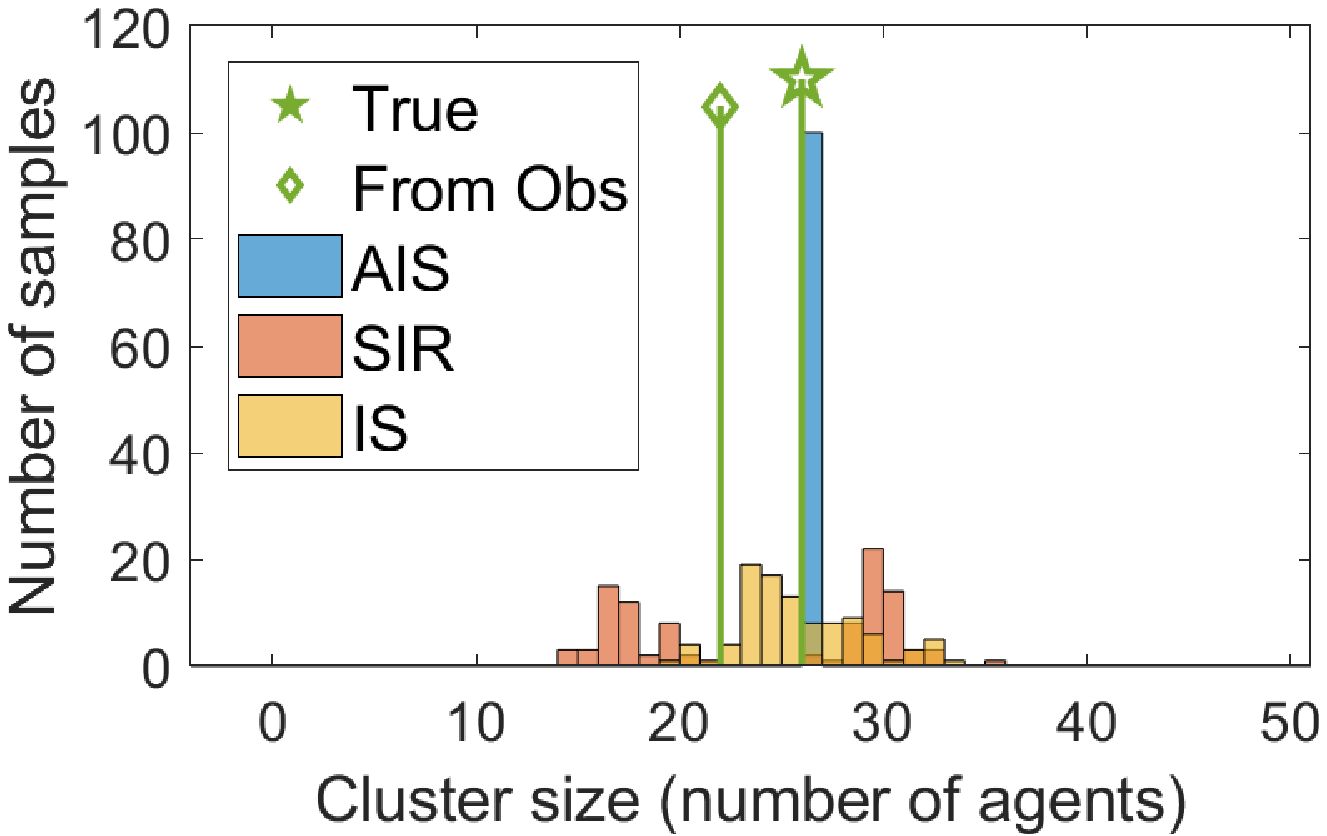}
		\hfil
		\includegraphics[width=0.485\linewidth]{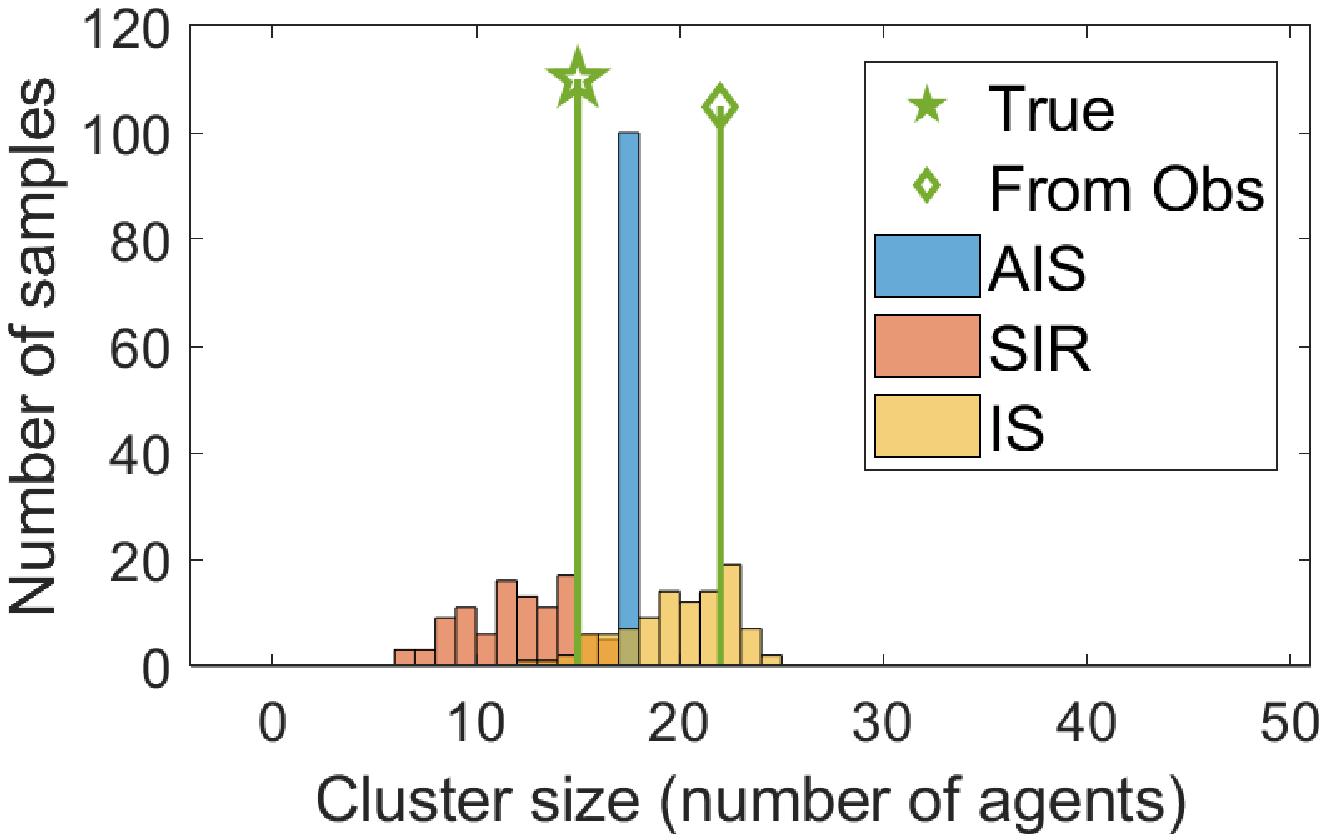}
		\hfil
		\caption{Posterior of cluster size: the  largest (left) and  2nd largest (right) cluster}
		\label{ToShow2ClusterNum0}
	\end{subfigure} 
	\quad
	\begin{subfigure}[b]{0.65\linewidth}
		\includegraphics[width=1\linewidth]{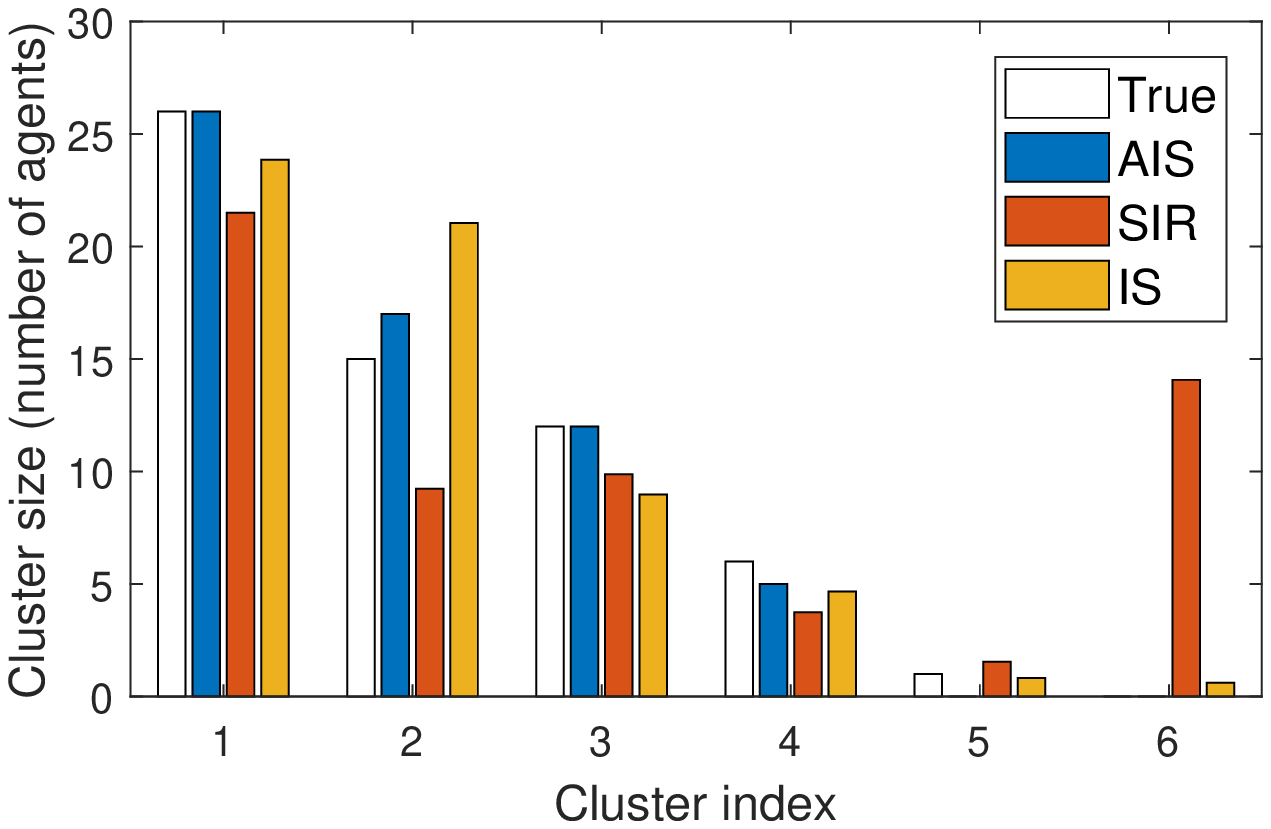}
		\caption{Sizes of all clusters (estimated by sample mean)}
		\label{Compare AIS with SIR0}
	\end{subfigure}
	\caption{Clustering prediction (noiseless observations): prediction of centers and sizes of clusters when observing $N_1=30$ of the $N=60$ agents for a short time. AIS algorithm outperforms SIR and IS at presenting concentrating around the truth posteriors of the centers and sizes for the leading clusters, and at providing accurate estimation of sizes for all clusters by sample mean. }   
	% In (\ref{ComparePolar0})-(\ref{Compare AIS with SIR0}), AIS algorithm (blue bars) outperforms SIR and IS. Look at the largest cluster: (\ref{ComparePolar0}) shows us that the difference between true center (white bar) and all samples from AIS (blue bars) is less than $0.1$ (bars overlapped), and a half samples from IS (yellow bars) also close to true center; (\ref{ToShow2ClusterNum0}) shows that all predicted cluster size from AIS is exactly the true size.}
	\label{Prediction for clusters noiseless}
\end{figure}

\paragraph{Noiseless observations}  Figure \ref{ComparePolar0}--\ref{ToShow2ClusterNum0}
show the empirical posteriors of the centers and sizes for the largest cluster and the second largest cluster (defined in Eq.\eqref{eq:postCluster}). Consider first the centers of the clusters in Figure \ref{ComparePolar0}. The true center of the largest cluster locates at the white bar in the left plot. All samples from AIS are close to the true center (with a distance less than $0.1$, resulting in a bar overlapping with the white bar of the true center). IS has about half samples at the true center and the other half far away from the true center. Nearly all samples of SIR mispredicted the true center.  Similar results can be seen in the right plot.

Consider next the cluster sizes in Figure \ref{ToShow2ClusterNum0}. The largest cluster has 27 agents, and the second largest cluster has 18 agents. The predicted sizes based on the observation $\x_T$ only (denoted by ``From Obs'') are both 22, not being able to identify the leading clusters. This suggests that the observation $\x_T$ itself is not enough to make an accurate prediction of the clustering. 
Among the SMC algorithms,  AIS leads to highly concentrated samples, at the true size for the largest cluster and near the true size for the second largest cluster. Implicit sampling (IS) leads to samples scattering around the truth. The samples of SIR scatter widely, tending to overestimate the size of the largest cluster and underestimate the size of the second largest cluster.  

Since the system is deterministic and the observations are noiseless, the true posterior of concentrates around the truth. AIS outperforms SIR and IS at representing such concentrated posterior.

Figure \ref{Compare AIS with SIR0} shows the sizes of all the clusters, estimated by sample mean as in \eqref{eq:mean}. AIS accurately captures the sizes of all the clusters except the smallest one, which is too small to be predicted. IS performed relatively well in predicting the largest cluster, but misses the 2nd largest cluster. SIR identifies the largest cluster with a relatively large error (5 relative to 27), but it misses all the other clusters.

%%%%%%%%%%%%%%%%
\begin{figure}[!t]
	\centering
	\begin{subfigure}[b]{1\linewidth}	\centering
		\includegraphics[width=0.48\linewidth]{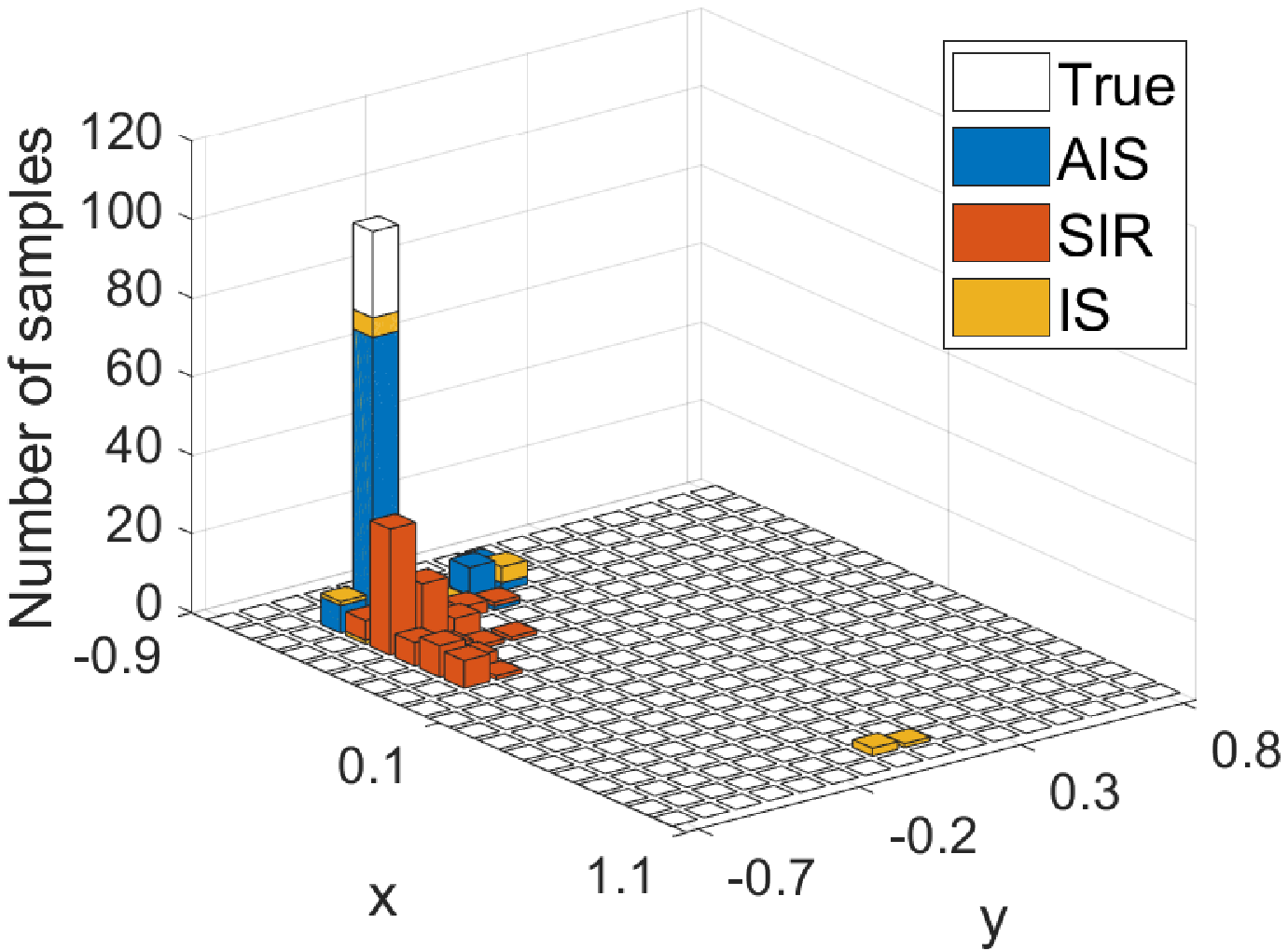}
\hfil
		\includegraphics[width=0.48\linewidth]{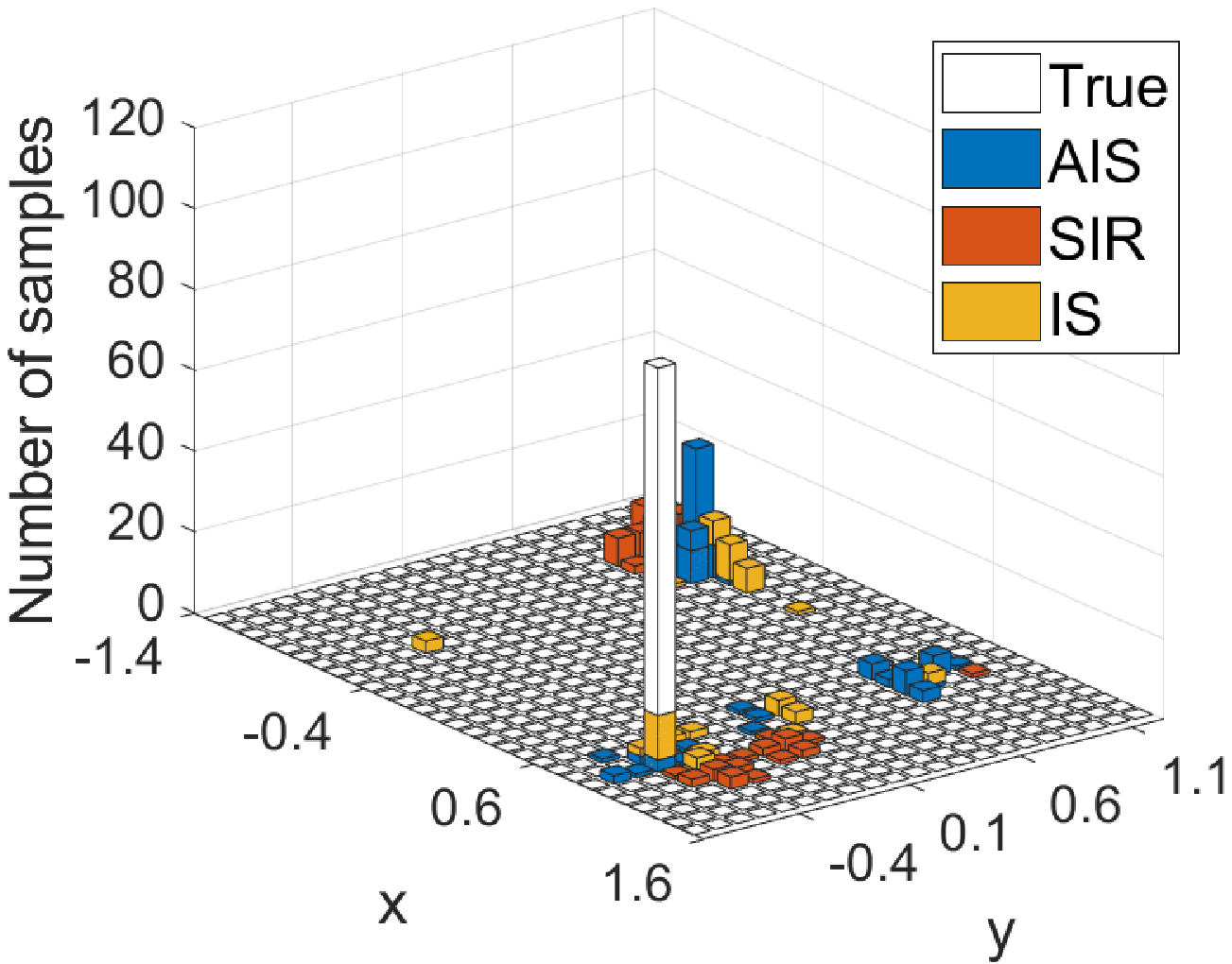}
\hfil
		\caption{Posterior of cluster center: the  largest (left) and 2nd largest (right) cluster}
		\label{ComparePolar}
	\end{subfigure} 
	\quad
	\vspace{.2cm}
	\begin{subfigure}[b]{1\linewidth}	\centering 
		\includegraphics[width=0.485\linewidth]{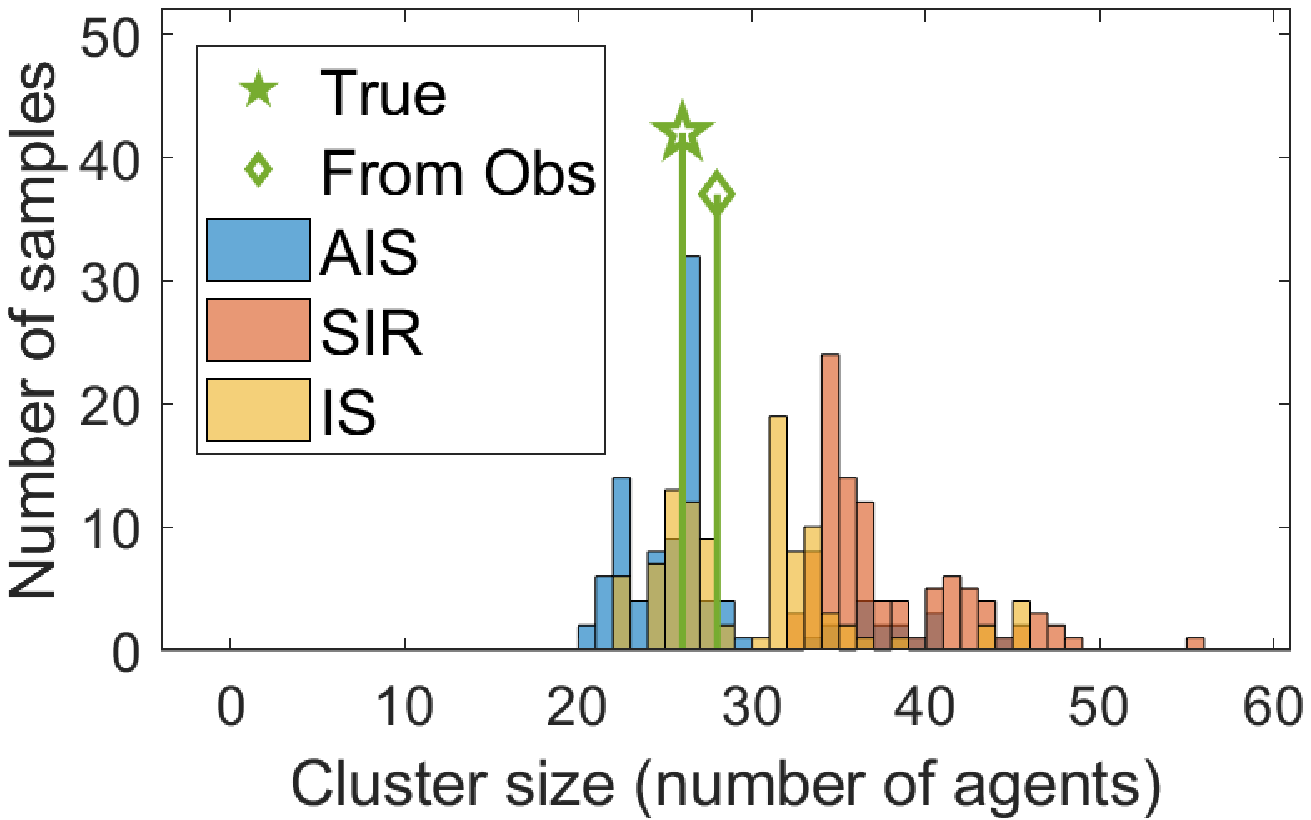}
\hfil
		\includegraphics[width=0.485\linewidth]{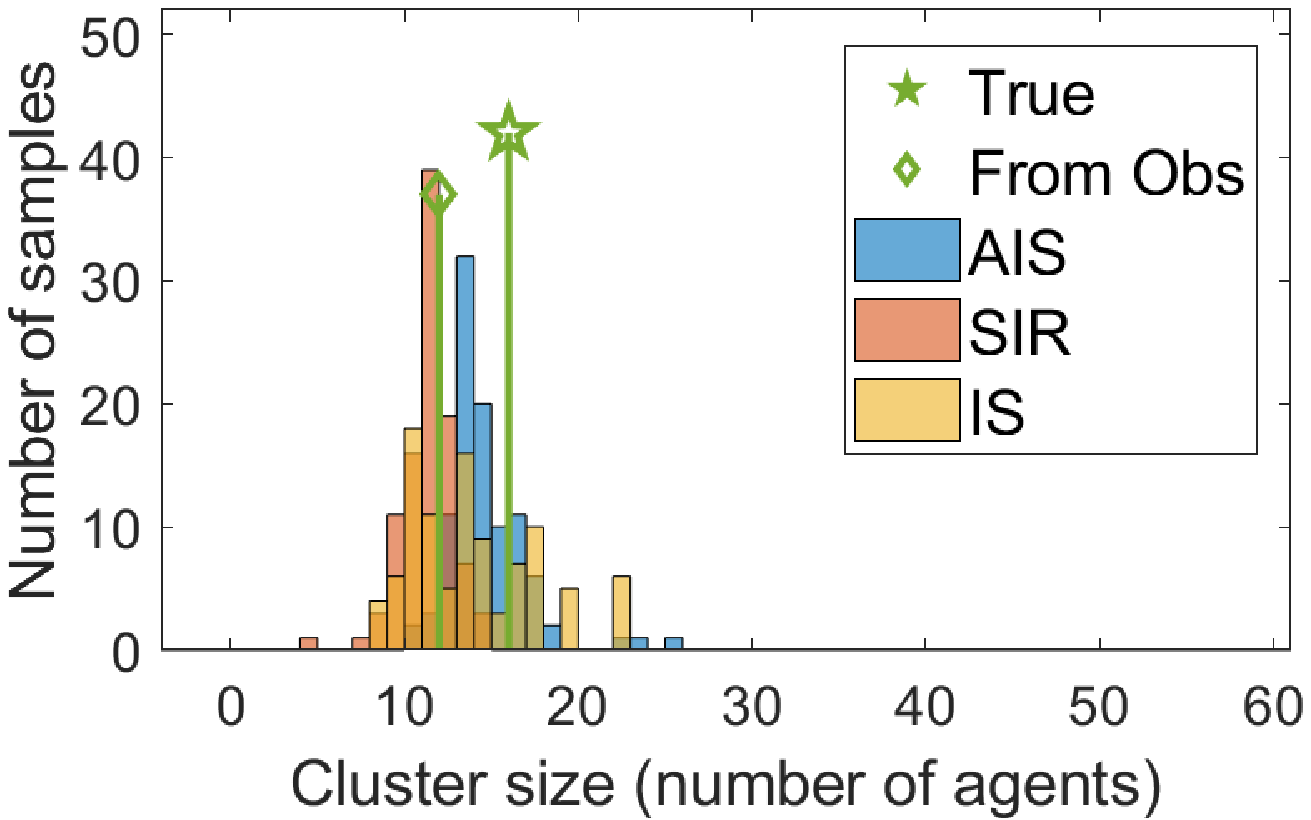}
\hfil
		\caption{Posterior of cluster size: the  largest (left) and 2nd largest (right) cluster}
		\label{ToShow2ClusterNum}
	\end{subfigure} 
	\quad
	\begin{subfigure}[b]{0.65\linewidth}
		\includegraphics[width=1\linewidth]{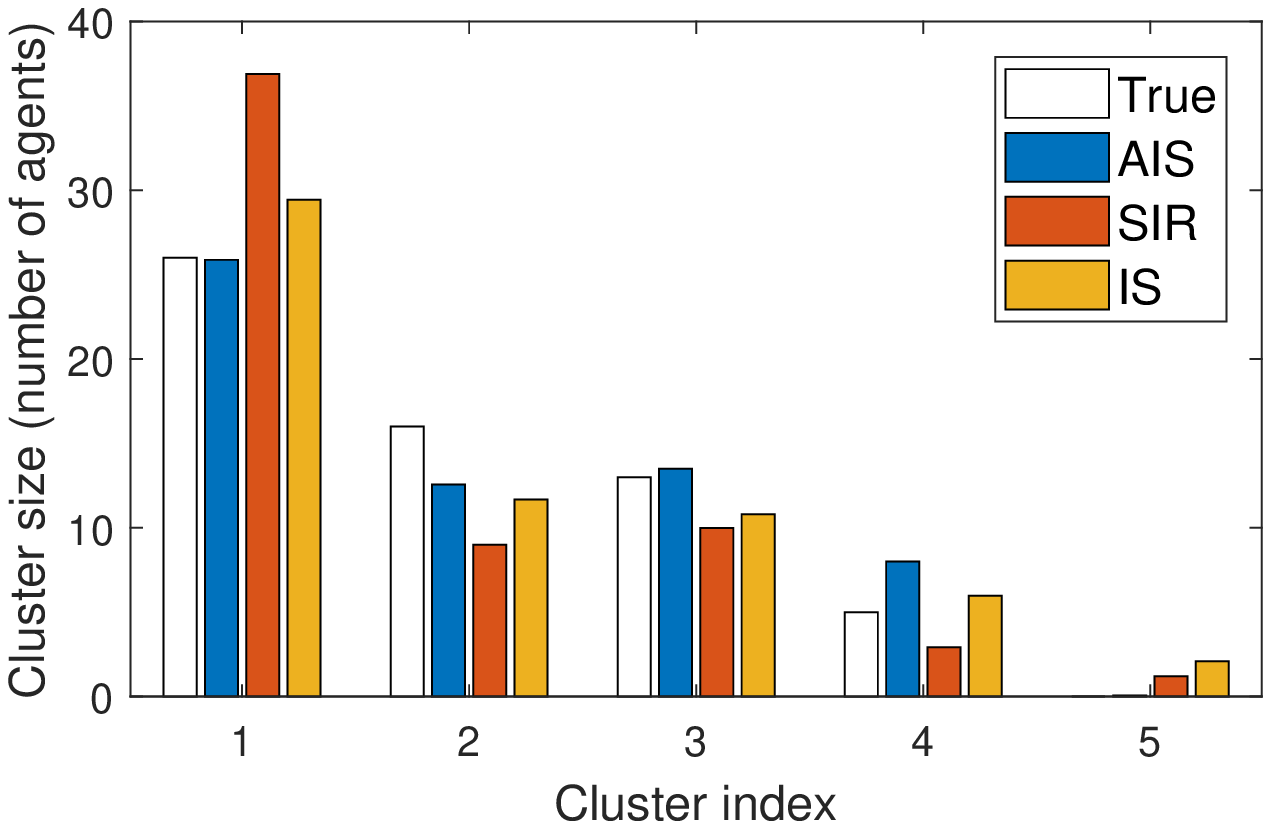}
		\caption{Sizes of all clusters (estimated by sample mean)}
		\label{Compare AIS with SIR}
	\end{subfigure}
	\caption{Clustering prediction (noisy observations): prediction of centers and sizes of clusters when observing $N_1=30$ of the $N=60$ agents for a short time with additive Gaussian noise. AIS algorithm performs slightly better than IS and clearly outperforms SIR,  at presenting posteriors of the centers and sizes for the leading clusters, and at providing accurate estimation of sizes for all clusters by sample mean. } 
	%Prediction for cluster center and size when observing $N_1=30$ of the $N=60$ agents with additive Gaussian noise. In (\ref{ComparePolar})-(\ref{Compare AIS with SIR}), AIS algorithm (blue bars) outperforms both SIR and IS. Result for largest cluster is similar to noiseless case; while (\ref{ComparePolar}) shows that both AIS (blue bars), SIR (red bars) and IS (yellow bars) failed to predict the correct center position (while bar).} 
	\label{Prediction for clusters noisy}
\end{figure}

\paragraph{Noisy observations} Figure \ref{Prediction for clusters noisy} shows the predictions when the observations are noisy. Due to the observation noise, the posteriors of the centers and sizes would present a larger uncertainty than the case of noiseless observations. Figure \ref{ComparePolar} shows the posteriors of centers for the two largest clusters. For the largest cluster, all three SMC algorithms yield samples scatter close to the true centers, with samples of AIS and IS concentrating at the true center more than those of SIR. For the second largest cluster, an unusual result appears: all the SMC algorithms lead to samples mostly missing the position of the true center. The reason is that the second and the third largest clusters have similar sizes (as shown in Figure \ref{Compare AIS with SIR}), the sizes are 16 and 13, respectively), causing difficulty in distinguishing them.

Figure \ref{ToShow2ClusterNum} presents posteriors of the sizes for the two largest clusters. The largest cluster has 26 agents, and the second largest cluster has 16 agents. AIS leads to samples concentrating at the true size for the largest cluster and near the true size for the second largest cluster. IS leads to samples scattering slightly wider than AIS, but are still around the truth. The samples of SIR scatter widely, tending to overestimate the size of the largest cluster and underestimate the size of the second largest cluster.  The predicted sizes based on the observation $\x_T$ only (denoted by ``From Obs'') are 28 and 12, slightly overestimating the size of the largest cluster and underestimating the size of the second largest cluster. 

Figure \ref{Compare AIS with SIR} shows the sizes of all the clusters, estimated by sample mean as in \eqref{eq:mean}. Only AIS accurately captures the size of the largest cluster, IS slightly overestimates the size, and SIR has an estimation that is too large. All three algorithms are able to lead to similar sizes for the second and the third clusters, with AIS being the closest to the truth. Note that all the SMC algorithms predict an untrue fifth cluster, but AIS has the smallest error.  

% \bigskip
In summary, for predicting centers and sizes of clusters from either noiseless or noisy observations, the performance of AIS algorithm is much better than that of SIR, which is usually not satisfactory, and is better than that of IS, which is often reasonably good.

%%%%%%%%%%%%%%%%==============
%%%%%%%%%%%%%%%%==============
%%%%%%%%%%%%%%%%==============
\subsection{Clustering prediction: success rates in many simulations} \label{Cp100}
We further investigate the clustering prediction by Auxiliary Implicit Sampling (AIS) in 100 independent simulations. We consider three cases: observing  $\frac{1}{2}$, $\frac{1}{3}$, and $\frac{1}{6}$ of the $60$ agents. We assess the performance by studying the success rate in predicting the centers for the largest two clusters, and the distribution of errors in size estimation. 
\paragraph{Assessment of the prediction performance} Recall that we estimate the centers and sizes of clusters by their posterior means. More precisely, the center $ \overline{\x}_{\C_i} $ and size $|\C_i|$ of cluster $\C_i$ are estimated by their sample means $ \widehat{\overline{\x}_{\C_i}}$ and $ \widehat{|\C_i|} $ as defined by \eqref{eq:mean}.  We denote by $\C_1$ and $\C_2$ the largest and the second largest clusters.

For each simulation, we say the center of the largest cluster $\C_1$ is predicted successfully if there exists an estimated cluster with a size in $[|\C_1|-K, |\C_1|+K]$ and with a center such that $\mathrm{dist}( \overline \x_{\C_1},  \widehat{\overline{\x}_{\C_j}}) <L$.  Here $K$ and $L$ are the levels of error tolerance. More specifically, we define an indicator function for a successful prediction of $\C_1$ by 
\begin{equation} \label{Cor function}
\Cor_1 = \left\lbrace \begin{aligned}
& 1,  \ \ \ \ \ \ \ \text{if } \mathrm{dist}\left(\overline{\x}_{\C_1}, \widehat{\overline{\x}_{\C_j}} \right) \leq L \text{ for some $j$ such that} 
\\
&  \ \ \ \ \ \ \ \ \ \ \ |\C_1|-K< \widehat{ |\C_j|} < |\C_1|+K; 
\\
& 0, \ \ \ \ \ \ \ \ \text{otherwise}, 
\end{aligned} \right. 
\end{equation}
In following simulations, we pick $L = 0.1$ (in general $L$ should depend on the communication function $\phi$, recall that our $\phi$ is supported in $[0,1]$) and the range of the value $K \in \{ 0,1,2 \}$. Similarly, we define a successful prediction for the center of the second largest cluster and its indicator function $ \Cor_2 $.

We access the prediction of the sizes of the largest two clusters by the distribution of the absolute error: 
\begin{equation} \label{evaluation function 1_2}
e_i =  \left | |\C_i^0| - \widehat{|\C_{i}|} \right|, \text{ for } i=1,2. 
\end{equation}
The error $e_i$ should be close to zero in a successful prediction. A heavy tail in the distribution of $e_i$ would indicate that it is difficult to predict the cluster size accurately.

\begin{figure}[!t]
	\centering
	\begin{subfigure}[b]{0.47\linewidth} 
		\includegraphics[width=1\linewidth]{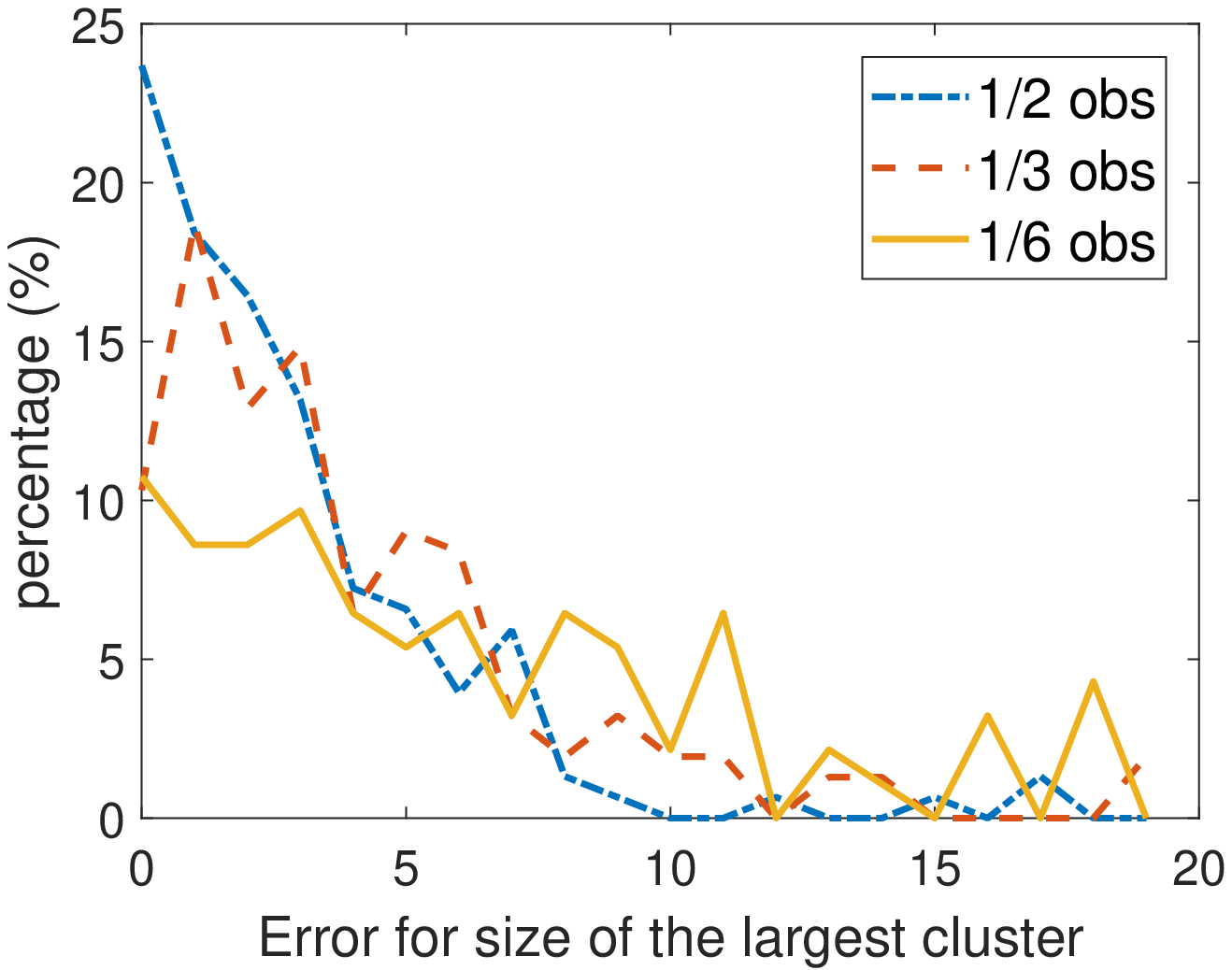}
		\caption{Absolute error of predicted size of the largest cluster}
		\label{errorbar1} 
	\end{subfigure}
	\hfil
	\vspace{.2cm}
	\begin{subfigure}[b]{0.47\linewidth} 
		\includegraphics[width=1\linewidth]{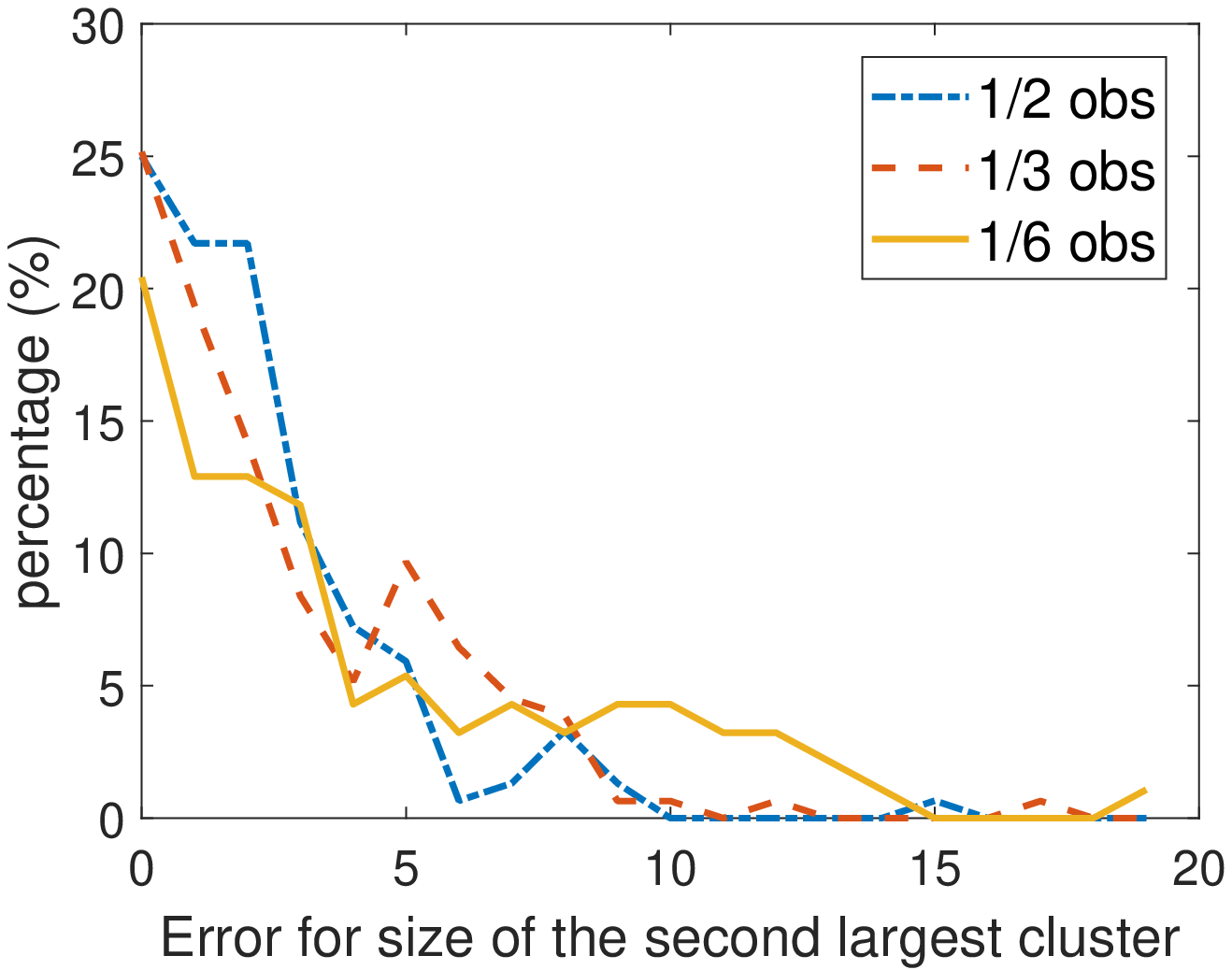}
		\caption{Absolute error of predicted size of the second largest cluster}
		\label{errorbar2}
	\end{subfigure}
	\hfil
	\quad
	\begin{subfigure}[b]{0.47\linewidth}
		\includegraphics[width=1\linewidth]{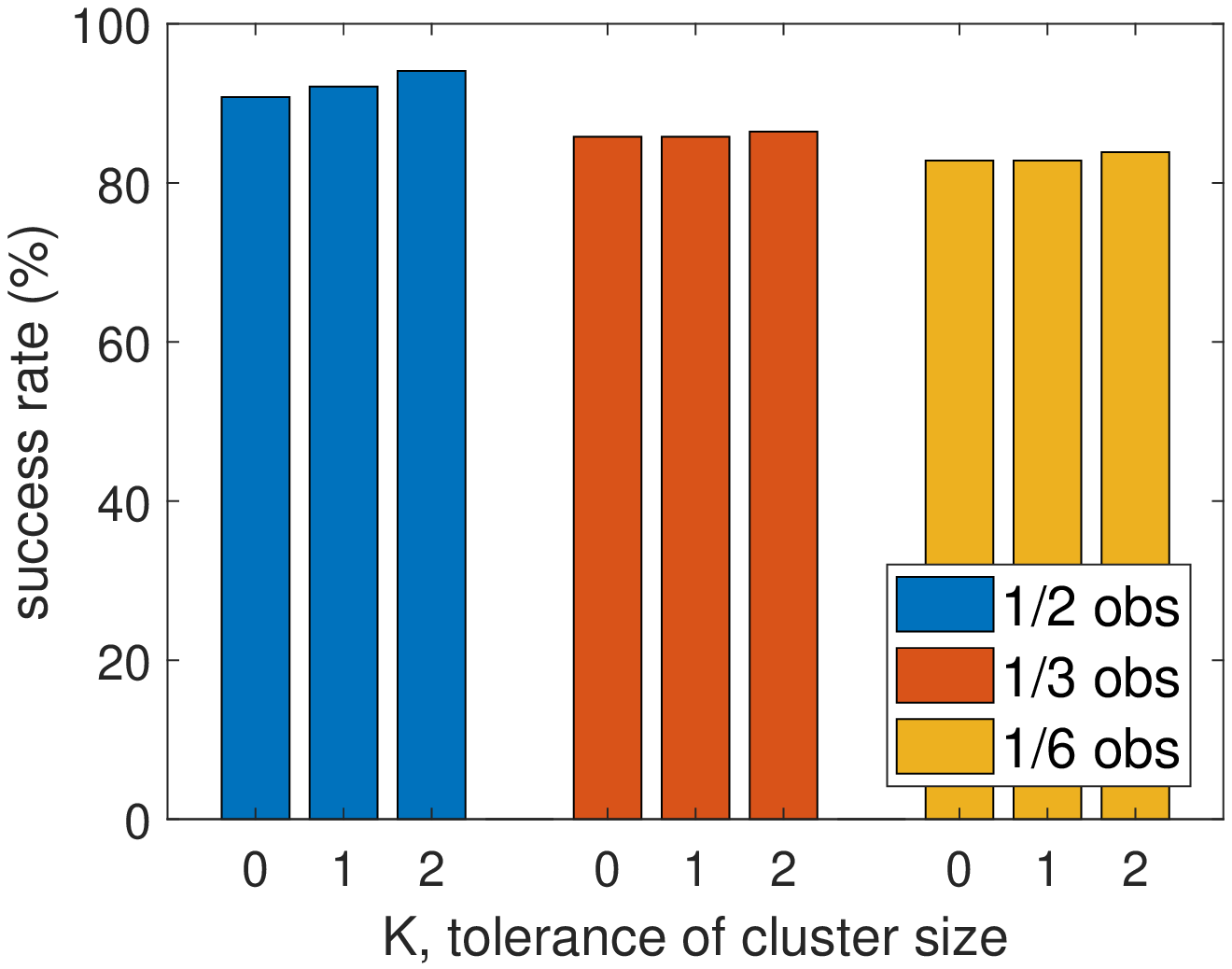}
		\caption{Center of the largest cluster}
		\label{CORRECT1}
	\end{subfigure}
	\hfil
	\begin{subfigure}[b]{0.47\linewidth}
		\includegraphics[width=1\linewidth]{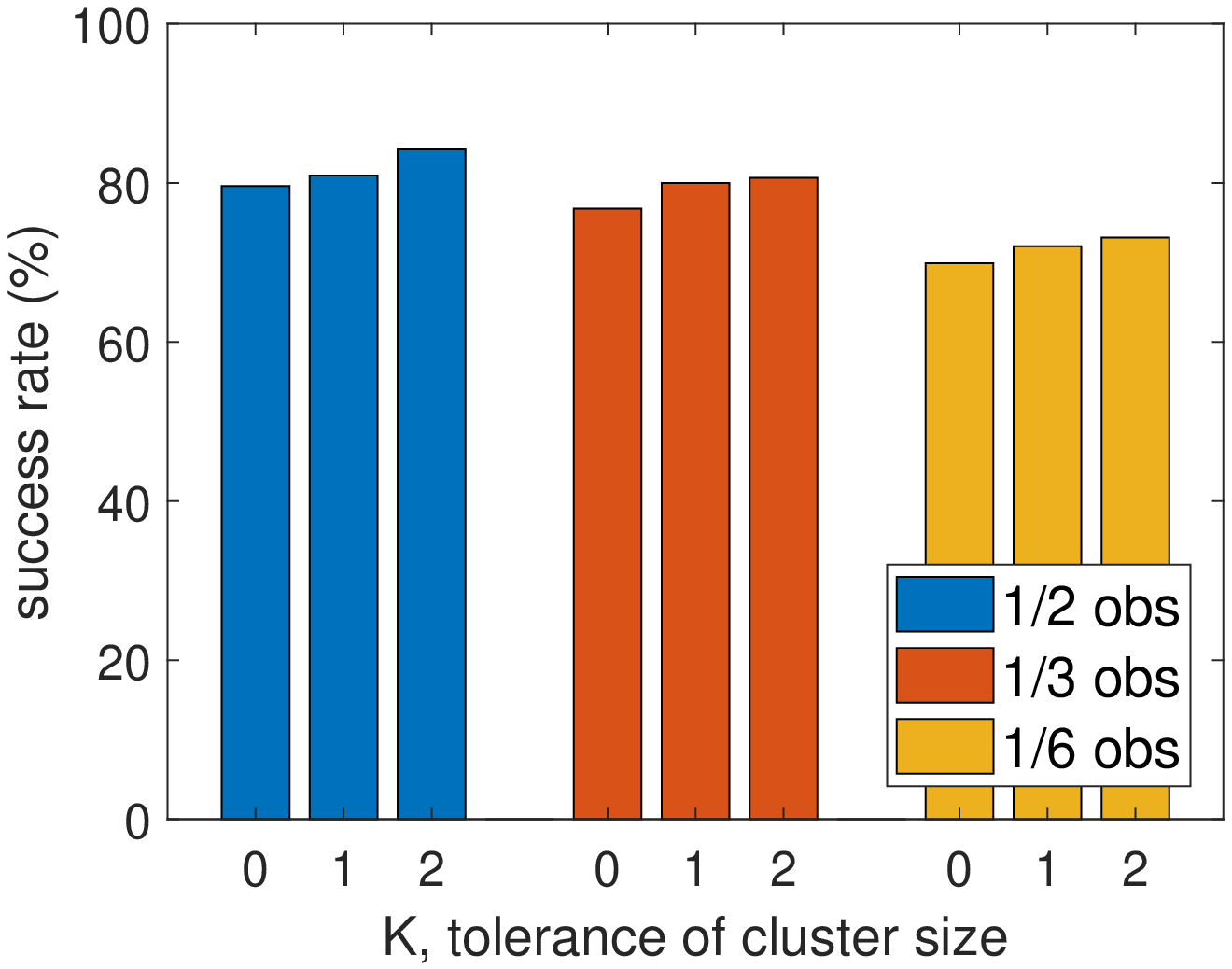}
		\caption{Center of the 2nd largest cluster}
		\label{CORRECT2}
	\end{subfigure}
	\hfil
	\caption{Cluster prediction in 100 simulations (noiseless case): observing $\ff 1 2$, $\ff 1 3$, or $\ff 1 6$ of the $N=60$ agents in the system. In (\ref{errorbar1})-(\ref{errorbar2}), the majority simulations (more than 70\%) can predict the cluster sizes reasonably, holding an error within 4, when the observation ratio is ether $\frac{1}{2}$ or $\frac{1}{3}$; but when the observation ratio is $\ff 1 6$, many simulations have large errors. 	 
		Figure \ref{CORRECT1}-\ref{CORRECT2} show that the centers of the leading clusters can be located with high probability ($85\%$-$95\%$ for the largest cluster and $75\%$-$85\%$ for the second largest cluster) even only observing $\ff 1 6$ of all agents. The success rate depends little on the tolerance level $K$. In short, the observation ratio affects the prediction of cluster sizes, but not the cluster centers. 
	}
	\label{result100Simu_noiseless}
\end{figure}

\begin{figure}[!t]
	\centering
	\begin{subfigure}[b]{0.47\linewidth}
		\includegraphics[width=1\linewidth]{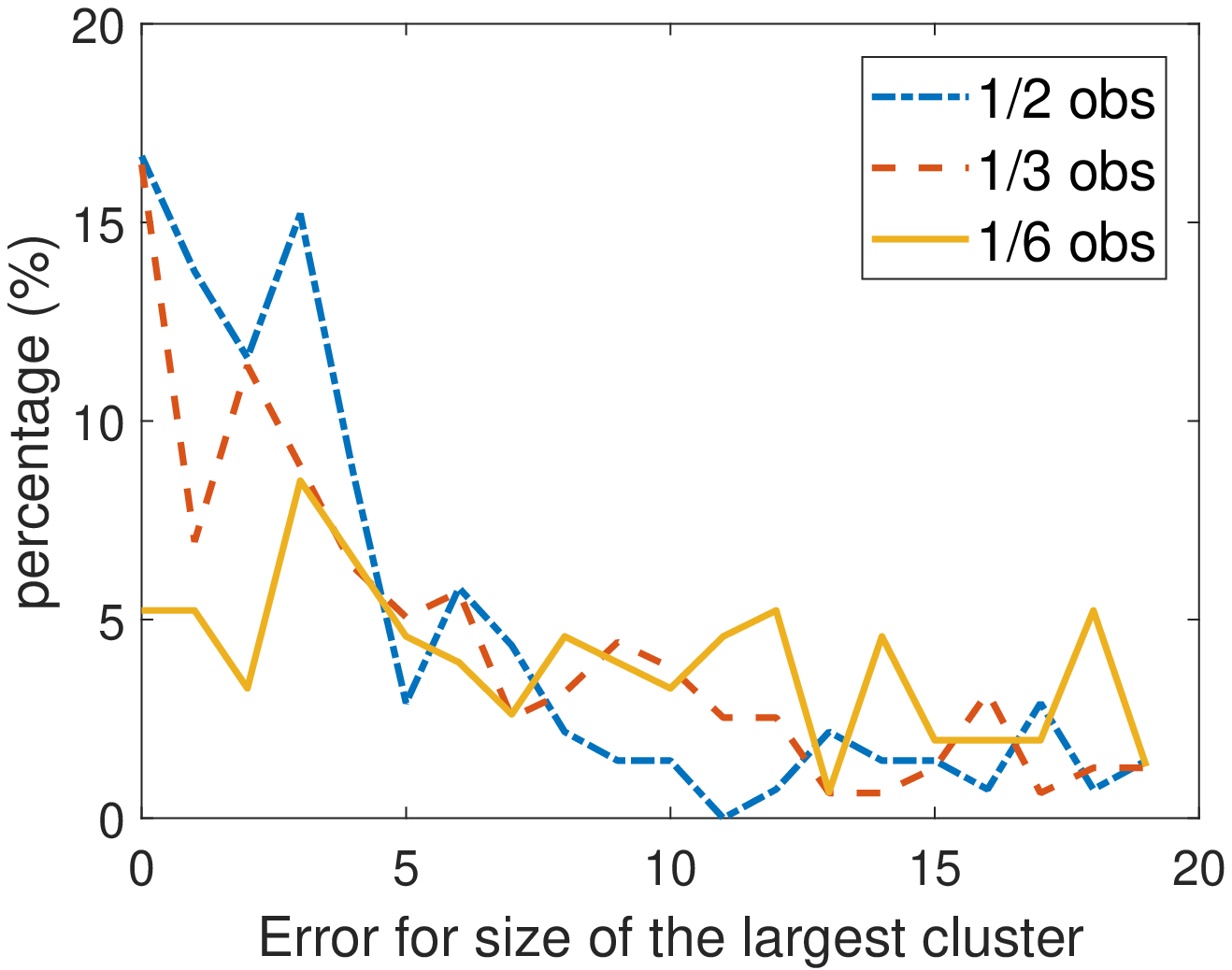}
		\caption{Absolute error of predicted size of the largest cluster}
		\label{errorbar3}
	\end{subfigure}
	\hfil
	\vspace{.2cm}
	\begin{subfigure}[b]{0.47\linewidth}
		\includegraphics[width=1\linewidth]{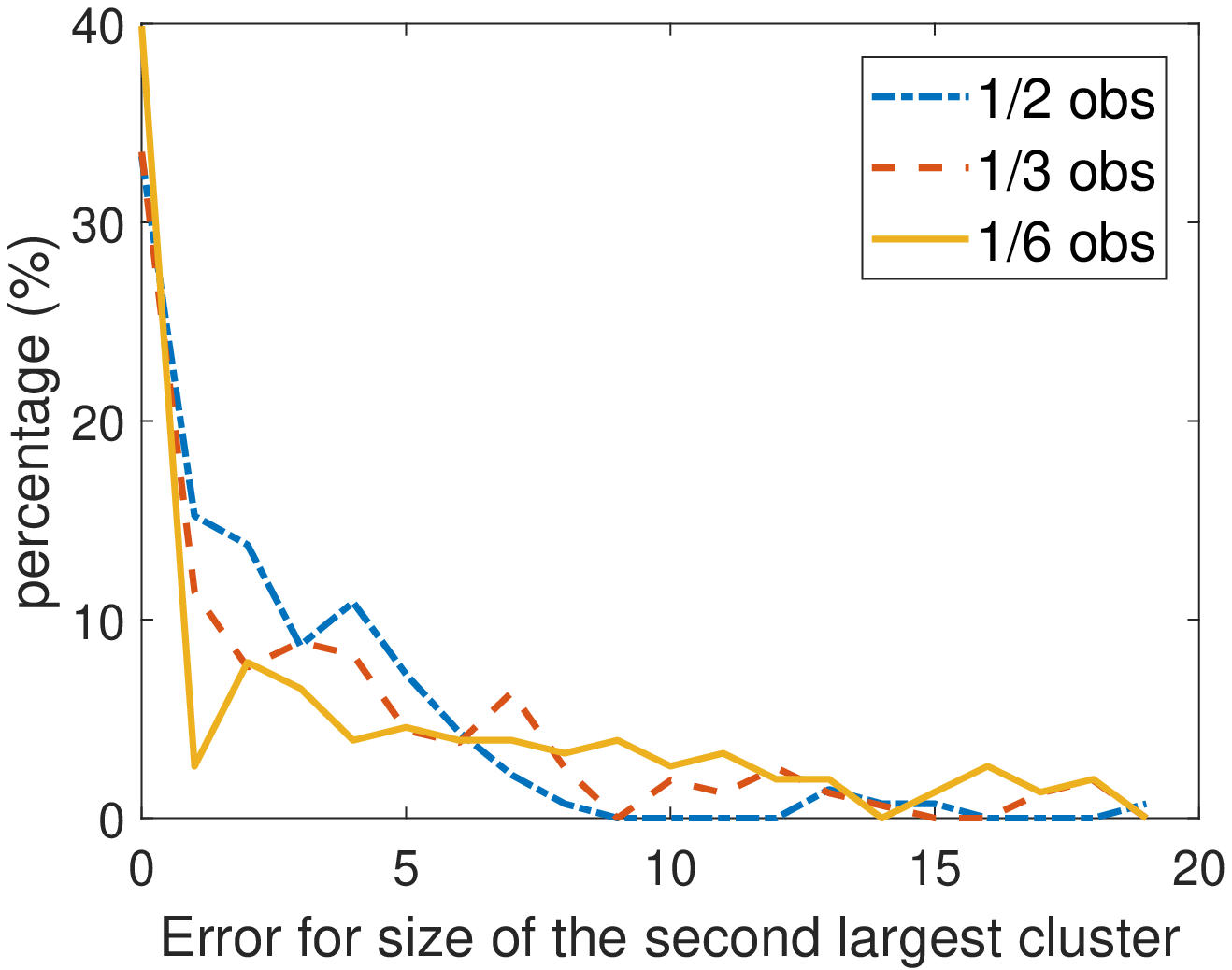}
		\caption{Absolute error of predicted size of the second largest cluster}
		\label{errorbar4}
	\end{subfigure}
	\hfil
	\quad
	\begin{subfigure}[b]{0.47\linewidth}
		\includegraphics[width=1\linewidth]{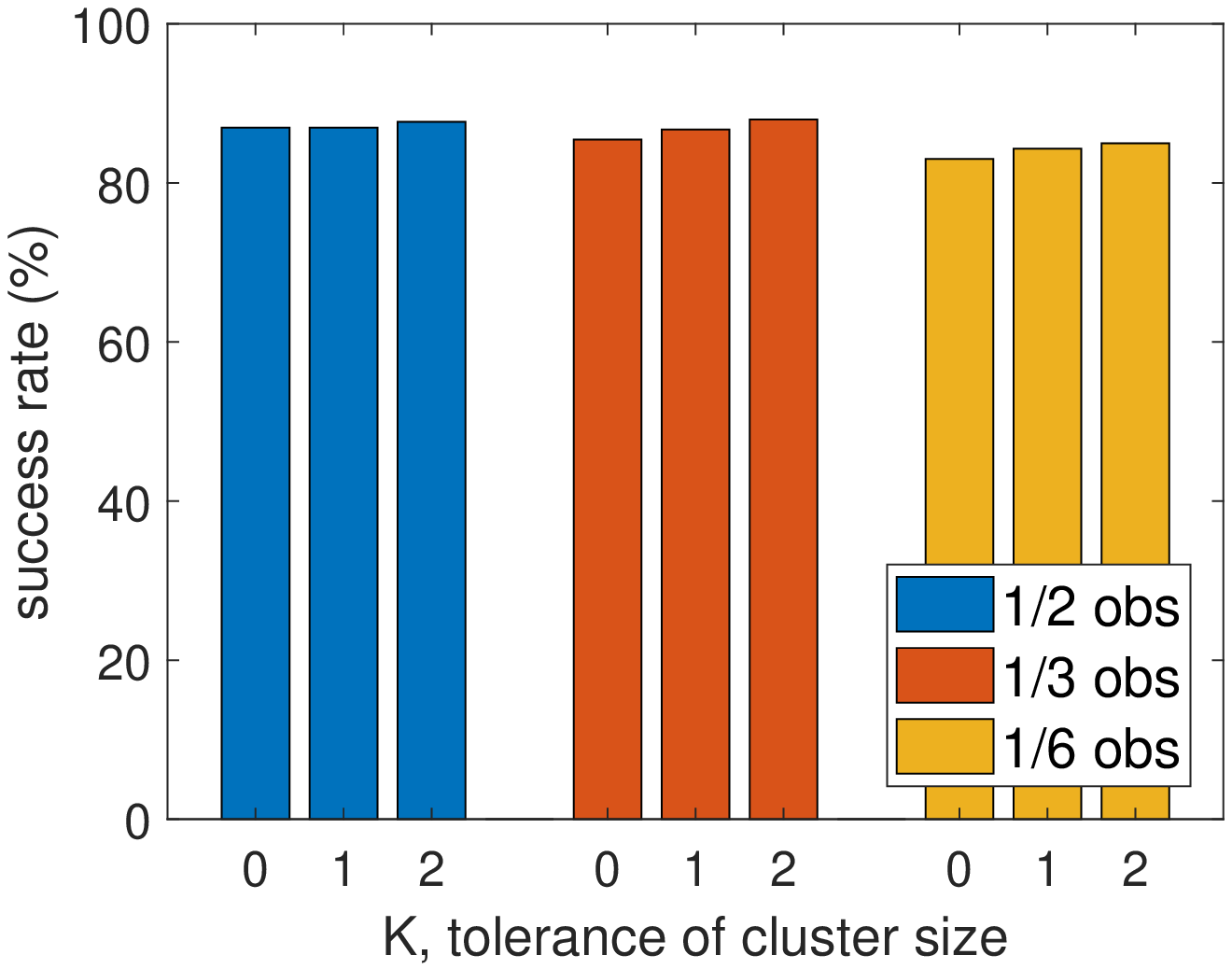}
		\caption{Center of the largest cluster}
		\label{CORRECT3}
	\end{subfigure}
	\hfil
	\begin{subfigure}[b]{0.47\linewidth}
		\includegraphics[width=1\linewidth]{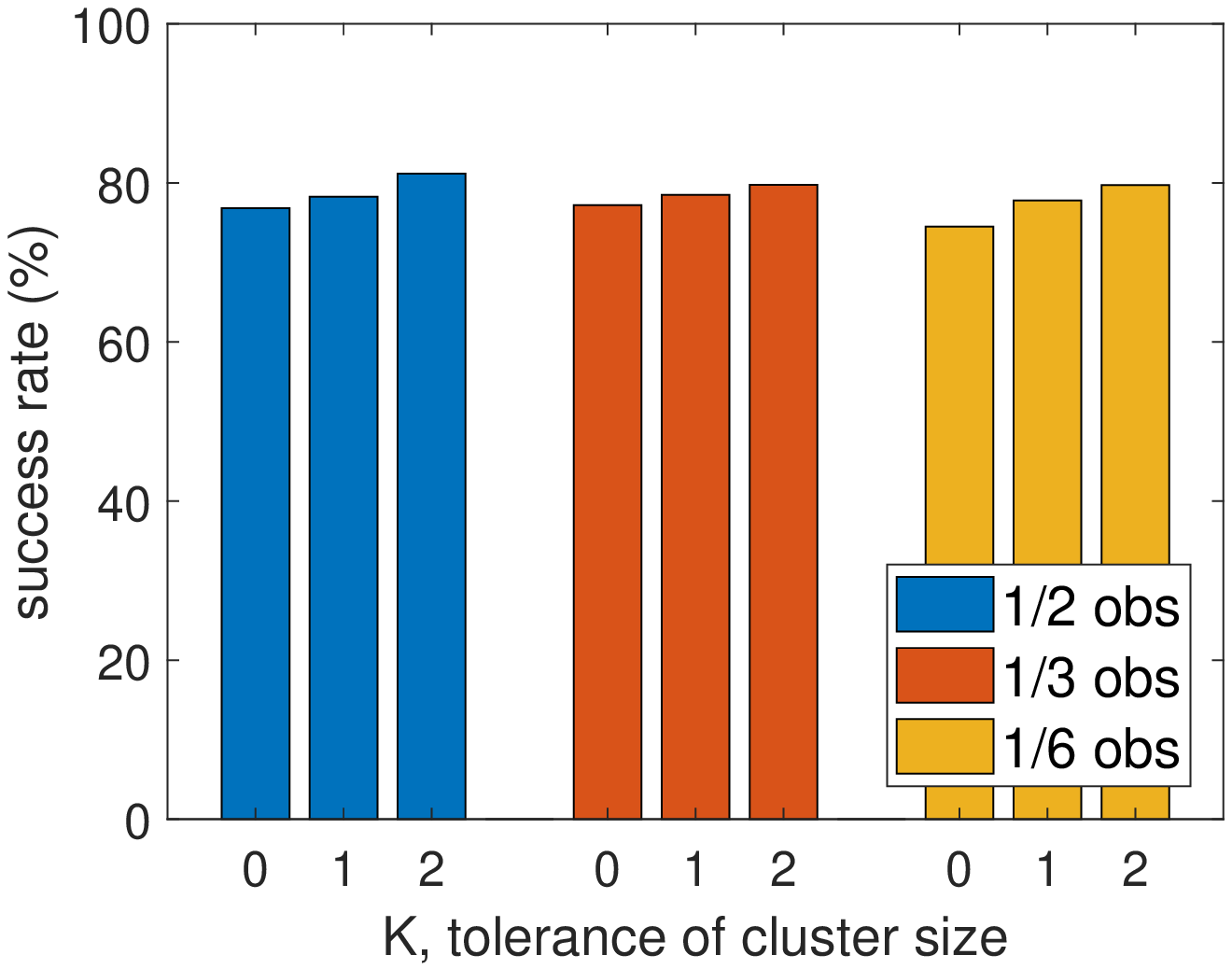}
		\caption{Center of the 2nd largest cluster}
		\label{CORRECT4}
	\end{subfigure}
	\hfil
	\caption{Cluster prediction in 100 simulations (noisy observations), observing $\ff 1 2$, $\ff 1 3$, or $\ff 1 6$ of the $N=60$ agents in the system with additive Gaussian noise. Similar to the case of noiseless observations: the prediction for sizes is more sensitive to observation ratio than the prediction of centers. In (\ref{errorbar3})-(\ref{errorbar4}), the error for size estimation is relatively large: when the observation ratio is $\ff 1 2$ or $\ff 1 3$, about $70\%$ of the simulations hold an error size less than 6; when the observation ratio is $\ff 1 6$, many simulations have large errors. In (\ref{CORRECT3})-(\ref{CORRECT4}), the centers are be predicted with a probability (around $85\%$ and $80\%$ for the largest and the second largest clusters, respectively), regardless of the observation ratio.}
	\label{result100Simu_noisy}
\end{figure}

\paragraph{Noiseless observations}
Figure \ref{result100Simu_noiseless} illustrates the performance of prediction of the largest two clusters in 100 independent simulations. We consider three observation ratios: $\ff 1 2$, $\ff 1 3$, or $\ff 1 6$, that is, observing $30$, $20$ and $10$ of the $N = 60$ agents in the system. The distributions of errors in the estimation of cluster sizes are shown in Figure \ref{errorbar1}-\ref{errorbar2}, and the success rate in predicting the centers are shown in Figure \ref{CORRECT1}-\ref{CORRECT2}.  

The prediction of cluster sizes depends on the observation ratio. When observing $\ff 1 2$ or $\ff 1 3$ of all agents, the majority simulations (more than 70\%) can predict the cluster sizes with an error within 4. But when the observation ratio is $\ff 1 6$, many simulations have large errors (larger than 4 for more than 50\% of the simulations), suggesting that the observations do not provide enough information for accurate prediction of the cluster sizes. 	 

The prediction of cluster centers exhibits a high success rate, regardless of the observation ratio. Figure \ref{CORRECT1}-\ref{CORRECT2} show that the centers of the leading clusters can be located with high probability $85\%$-$95\%$ for the largest cluster and $75\%$-$85\%$ for the second largest cluster, and that the successes rate drops slightly when the observation ration decreases from $\ff 1 2$ to $\ff 1 6$. Also, the success rate depends little on the tolerance level $K$.

\paragraph{Noisy observations} When the trajectories are observed with additive Gaussian noise, similar to the case of noiseless observations, the prediction for sizes is more sensitive to observation ratio than the prediction for centers. Due to the additional uncertainty from the observation noise, the error for size estimation is larger than the noiseless case. In (\ref{errorbar3})-(\ref{errorbar4}),  when the observation ratio is $\ff 1 2$ or $\ff 1 3$, about $70\%$ of the simulations hold an error size less than 6; when the observation ratio is $\ff 1 6$, about $50\%$ of the simulations have errors larger than $4$. In particular,  the size of the second largest cluster is predicted more accurately than the largest cluster, indicating that the observation noise is mostly absorbed in the prediction of the leading cluster. 

The observation noise also slightly reduces the success rate in the prediction of cluster centers. In (\ref{CORRECT3})-(\ref{CORRECT4}), the centers are predicted with a high probability (around $85\%$ and $80\%$ for the largest and the second largest clusters, respectively), regardless of the observation ratio.

In summary, for either noiseless or noisy observations, the cluster center can be predicted with a high success rate, regardless of the observation ratio. The cluster size, on the other hand, has a larger uncertainty that is sensitive to both the observation noise and the ratio of observation. %are more difficult to be predicted accurately, and is sensitive to the observation ratio.  

%%%%%%%%%%%%%%%%%==================
%%%%%%%%%%%%%%%%%==================
\section{Discussion and conclusion}\label{sec:conclusion}
We presented a Bayesian formulation for clustering prediction of opinion dynamics from partial observations, characterizing the prediction by the posterior of the clusters' sizes and centers. To overcome the challenge in sampling the high-dimensional posterior with multiple local maxima, we introduced an auxiliary implicit sampling (AIS) algorithm using two-step observations, which is a sequential Monte Carlo (SMC) method that combines the ideas from auxiliary particle filters \cite{pitt1999filtering} and implicit particle filters \cite{CT09}. In both cases of noiseless and noisy observations, the AIS algorithm leads to accurate predictions of the sizes and centers for the leading clusters. 

\cyan{The uncertainty in the posterior increases when the ratio of the observed population decreases. Remarkably, the cluster center can be predicted with a high success rate, regardless of the observation noise and ratio. This suggests that the centers have relatively small uncertainty, agreeing with the fact that they are the average of the agents' opinions. The cluster size, on the other hand, has a considerable uncertainty that is sensitive to both the observation noise and the ratio of observation. 
}

There are three directions for future research. First, improve the information in observation by a random selection of agents to observed at each time. The observations in this study are trajectories of a fixed set of agents, which may yield little information about other clusters when the observations concentrate in one cluster. Random selection of agents may avoid such an information loss by providing an unbiased sampling of all the agents' opinions. Second, \cyan{extension to large systems with millions of agents using mean-field equations. When there are millions of agents, it becomes computationally prohibitive to simulate the ODEs, and it is natural to consider the corresponding mean-field equation for the concentration density of the agents' opinions 
% One may describe the concentration density of the opinions by the mean-field equation 
(see e.g., \cite{motsch2014heterophilious,lang2020learning,bauso2016opinion}). 
   % apply sequential Monte Carlo methods for the mean-field equation 
Extension of our AIS method is straightforward. The major issue is the computational cost when solving the mean-field PDE many times, and one may have to use reduced models (see e.g.,\cite{CLMMT16,LTC17}) to achieve efficiency.} Third, learn both the states and the communication function or the network topology \cite{coutino2020_StateSpaceNetwork} from partial noisy observations, either for systems with finite agents or for the mean-field equation. \blue{Our AIS algorithm supplies the SMC part for algorithms that combines SMC with MCMC, such as the particle Gibbs methods or the nested particle filters  \cite{andrieu2010particle,crisan2018nested,lindsten2014particle}, to jointly estimate the parameters and states. }

\section*{Acknowledgment}
The authors would like to thank the three anonymous referees for valuable comments. The authors thank Mauro Maggioni and Sui Tang for inspiring discussions. The authors are grateful for supports from NSF-1913243, NSF-1821211, MARCC, and Johns Hopkins University. 

\ifCLASSOPTIONcaptionsoff
  \newpage
\fi


%\bibliographystyle{IEEEtran}
%\begin{thebibliography}{1}
%\end{thebibliography}


%\cite{IEEEexample:articleetal}
%
%\bibliographystyle{IEEEtran}
%\bibliography{IEEEabrv,IEEEexample}

\bibliographystyle{IEEEtran} 
\bibliography{IEEEabrv,Cite_191212,learning_dynamics,ref_FeiLU,OpinionDS_ref,MonteCarlo}


%\end{IEEEbiography}
%% insert where needed to balance the two columns on the last page with
%%\newpage


%\vfill


\end{document}